\documentclass[letterpaper,draft,onecolumn,11pt]{IEEEtran}

\renewcommand{\baselinestretch}{1.6}
\usepackage{amsmath,amssymb,bm}
\usepackage{epsfig,subfigure,colortbl,psfrag,subfloat}
\usepackage{enumerate,array,multirow}
\usepackage{cite,url}
\usepackage{algorithmic}
\usepackage{algorithm}
\usepackage{stfloats}
\usepackage{xcolor}
\newtheorem{remark}{Remark}

\def \diag{\operatornamewithlimits{diag}}

\def \tr{\operatornamewithlimits{tr}}

\def \minj{\operatornamewithlimits{min}}
\def \maxj{\operatornamewithlimits{max}}

\def \log{\operatorname{log}}

\def \arg{\operatorname{arg}}

\def \E{\operatorname{E}}

\def \dB{\operatorname{dB}}

\def \dim{\operatornamewithlimits{dim}}

\def \dB{\operatorname{dB}}
\def \dBm{\operatorname{dBm}}

\def \GHz{\operatorname{GHz}}
\def \uW{\operatorname{\mu W}}
\def \nW{\operatorname{nW}}
\def \W{\operatorname{W}}
\def \m{\operatorname{m}}
\def \km{\operatorname{km}}

\def \us{\operatorname{\mu s}}
\def \ns{\operatorname{ns}}

\def \bpsHz{\operatorname{b/s/Hz}}

\def \Mb{\operatorname{Mbps}}
\def \Gb{\operatorname{Gbps}}

\def \Kmph{\operatorname{km/h}}
\def \bpJ{\operatorname{b/J}}

\def \MbpJ{\operatorname{Mb/J}}

\def \pWpbps{\operatorname{pW/bit/s}}
\def \MHz{\operatorname{MHz}}
\def \Hz{\operatorname{Hz}}
\def \GHz{\operatorname{GHz}}

\def \beq {\begin{equation} }
\def \eeq {\end{equation} }
\def \beqn{\begin{eqnarray} }
\def \eeqn{\end{eqnarray} }
\def \bmat{\begin{bmatrix}}
\def \emat{\end{bmatrix}}
\def \bmats{\left(\begin{smallmatrix}}
\def \emats{\end{smallmatrix}\right)}
\def \beqi{\begin{IEEEeqnarray}{rcl}\IEEEyesnumber}
\def \eeqi{\end{IEEEeqnarray}}
\def \inum{\IEEEyessubnumber}

\newcommand{\Pfix}{P_{\rm fix}}

\psfull

\begin{document}

\pagestyle{empty}

%

\title{Energy-Efficient, Large-scale Distributed-Antenna System (L-DAS) for Multiple Users}
\author{Jingon~Joung$^{*}$,~\IEEEmembership{Member,~IEEE,}
        Yeow~Khiang~Chia,~\IEEEmembership{Member,~IEEE,}
        and~Sumei~Sun,~\IEEEmembership{Senior~Member,~IEEE}
%
%
\thanks{The authors are with the Institute for Infocomm Research (I$^2$R), A$^\star$STAR, Singapore 138632; $^*$J. Joung (e-mail: jgjoung@i2r.a-star.edu.sg) is the corresponding author.}
}
\markboth{IEEE JOURNAL OF SELECTED TOPICS IN SIGNAL PROCESSING, VOL. 00, NO. 0, MONTH 2013}%
{Joung \MakeLowercase{\textit{et al.}}: Energy-Efficient, Large-scale Distributed-Antenna System (L-DAS) for Multiple Users}
\IEEEpubid{1932-4553/\$31.00~\copyright~2013 IEEE}
\maketitle
\pagestyle{empty}
\thispagestyle{empty}

\begin{abstract}
Large-scale distributed-antenna system (L-DAS) with very large number of distributed antennas, possibly up to a few hundred antennas, is considered.  {A few major issues of the L-DAS, such as high latency, energy consumption, computational complexity, and large feedback (signaling) overhead, are identified.} The potential capability of the L-DAS is illuminated in terms of an {\it energy efficiency} (EE) throughout the paper. We firstly and generally model the power consumption of an L-DAS, and formulate an EE maximization problem.
 {To tackle two crucial issues, namely the huge computational complexity and large amount of feedback (signaling) information, we propose a channel-gain-based antenna selection (AS) method and an interference-based user clustering (UC) method. The original problem is then split into multiple subproblems by a cluster, and each cluster's precoding and power control are managed in parallel for high EE. Simulation results reveal that i) using all antennas for zero-forcing multiuser multiple-input multiple-output (MU-MIMO) is energy inefficient if there is nonnegligible overhead power consumption on MU-MIMO processing, and ii) increasing the number of antennas does not necessarily result in a high EE. Furthermore, the results validate and underpin the EE merit of the proposed L-DAS complied with the AS, UC, precoding, and power control by comparing with non-clustering L-DAS and colocated antenna systems.}
\end{abstract}


\section{Introduction}\label{Sec:Intro}


\PARstart{L}{arge-scale} (or massive) multiple-input multiple-output (L-MIMO) techniques have been rigorously studied to tremendously improve spectral efficiency (SE, $\bpsHz$) of wireless communications. The L-MIMO employs very large number of colocated antennas, which can effectively mitigate {\it small-scale} (local) distortion such as noise at the receiver and fast fading ( {see \cite{NgLoSc12a,BjHoKoDe13,BjSaHoDe13,RuPeLaLaMaEdTu13}} and the references therein). 
On the other hand, a distributed-antenna system (DAS) is also one of the promising technologies to effectively improve SE of wireless communications.  {DAS is implemented with multiple distributed antennas (DAs) through base stations (BSs) located in different cells, i.e., coordinated multipoint (CoMP) transmission (see \cite{ChPePh13} and references therein)}, or through distributed radio remote heads typically located in the same cell. DAS can mitigate {\it large-scale} fading (path loss) using many antennas distributed geographically. The SE of DAS has been mainly studied (see. e.g., \cite{DaZhYa05,KiLeLeLe12,ChAn07,ZhAn08,HePeWaZh13}). In \cite{DaZhYa05}, the authors show proportional relationship between SE and the number of DAs. A suboptimal power control method and a simple antenna selection (AS) method are proposed to improve the SE \cite{KiLeLeLe12}. It is shown that a single DA usage is preferable to full DA usage in multicell scenario \cite{ChAn07}, while the opposite results are observed in a single, isolated cell \cite{ZhAn08} and also in a multiuser (MU) scenario \cite{HePeWaZh13}.


DAS's channels are typically modeled as the composite channels including {\it uncorrelated} large- and small-scale fading channels, which are a crucial part of motivation of a DAS technique and differentiate the DAS from L-MIMO techniques. Note that all colocated antennas suffer almost {\it identical} large-scale fading and {\it highly correlated} small scale fading.  {With the common goal, i.e., high SE, it is natural step to consider very {\it large-scale DAS} (L-DAS) to obtain the synergy for further SE improvement}. Nevertheless, the L-DAS has rarely been studied due to the ambiguity of cost of {\it very large-size} networks. To observe the tradeoff between the cost and benefit, and to quantify the efficacy of L-DAS, we consider an energy efficiency (EE, $\bpJ$) that is the total amount of reliably decoded bits normalized by the consumed energy. The EE is a widely used metric in wireless communications recently to find a Pareto optimality between throughout and energy consumption (see e.g., 
\cite{JoHoSu13WCL,JoHoSu14JSAC}). The SE-EE tradeoff has been analyzed for a single user (SU) in DAS systems \cite{OnHeIm13}, and EE optimal power control has been proposed for DAS to support an SU \cite{KiLeSoLe13}. 
In \cite{CaLiZhWa07}, the authors show convergence of total transmit power and per-user sum rate when the number of DAs and users approach to {\it infinity} like an L-DAS, and provide an asymptotical EE that is a ratio between per-user sum rate and total transmit power. However, total transmit power does not imply total energy consumption in practice due to the overhead energy consumption at the transmitter. Though the overhead to process additional DAs has been recently addressed in \cite{CaGa10}, the EE behavior of L-DAS is still unclear and difficult to be conjectured from the existing studies, especially for an MU scenario. Note that the power consumption of L-DAS prohibitively increases as the network size increases.

\pubidadjcol

In this paper, the EE of L-DAS is studied in MU scenario. We model an L-DAS transmitter and its power consumption including the overhead. An EE maximization problem is formulated under constraints on per-antenna transmit power and per-user rate to select DAs, design MU precoding, and control transmit power. To tackle the original, computationally intractable optimization problem, we propose a channel-gain-based AS method and a interference-based user clustering (UC) method, which enable us to split the original problem into multiple cluster-based subproblems. The multiple subproblems can be solved in parallel, resulting in computational complexity and feedback (signaling) reduction. Each subproblem is further divided into precoding design and power control problems. {EE-aware precoding is derived for each cluster, and the per-cluster optimal and heuristic power control algorithms are then proposed.} To further improve EE, additional DA assignment and clustering threshold adaptation are considered. Simulation with practical parameters is performed to observe an average EE over clustering threshold, number of users, and network size. The results of the paper will be a useful reference for further study of energy efficient L-DAS. The summary of main contributions and results of our work:
%
%
 {\begin{itemize}
    \item We propose an L-DAS, which is a new, natural extension of L-MIMO to DAS systems.
    \item We provide a practical power consumption model for the L-DAS, which can be readily modified and applied to any types of distributed systems (Section \ref{Sec:ProbForm}).
    \item We formulate an EE maximization problem for a general L-DAS setup (Section \ref{Sec:ProbForm}), and solve it through a suboptimal, decomposition strategy.
    \item We propose simple AS and UC methods to split the original problem (Section \ref{Sec:AntSel}), which enables a cluster-based design, resulting in reduction of computational complexity and signaling overhead.
    \item We generalize the results in \cite{JoChSu13} for the precoding (Section \ref{Sec:ZF}) and per-cluster optimal and heuristic power control (Section \ref{SEC:Pow}) of L-DAS.
    \item We show informative simulation results, under the practical power consumption model (Section \ref {Sec:Eval}).
    \begin{itemize}
    \item Using all DAs for MU-MIMO could be an energy-inefficient strategy if there is nonnegligible overhead power consumption for MU-MIMO precoding.
    \item Increasing the number of DAs does not necessarily result in a high EE. In other words, there exists EE optimal network size.
    \item The proposed L-DAS complied with the AS, UC, precoding, and power control improves EE compared to non-clustering L-DAS and colocated antenna systems.
        \end{itemize}
\end{itemize}}


\section{L-DAS System  {and Its Issues}}\label{Sec:Syst}
We consider an L-DAS with one central unit, called a baseband unit (BBU) or a signal processing center, $M$ DAs, and $U$ user equipments (UEs) (refer to Fig. 2(a) in Subsection \ref{Sec:CU}). L-DAS has a very large number of DAs compared to the number of UEs, i.e., $M\gg U$. For simple demonstration, a grid antenna layout is depicted in Fig. 2(a), yet any type of antenna layouts, such as circular and random layouts, can be applied to our system model\footnote{A generalized DAS employing multiple distributed transmitters with multiple colocated antennas,  {e.g., CoMP \cite{ToPeKo11,NgLoSc12b},} or employing multiple DAs \cite{WaZhCh09,JoSu13,JoChSu13} can be exploited as the L-DAS. Note that multiple transmitter can be implemented by multiple DAs with identical PAs, which is a reasonable approach based on the results in \cite{JoSu13}, which reports that {\it equal} power output capability (POC) of transmitters provides further EE merit compared to the unequal POC.}.  {All DA ports are connected to the BBU through a noise-free wired fronthaul for coordinated and cooperative communications. Since a passive optical network (PON) can support data rate up to $2.4\Gb$ with low power consumption of around $1\W$ per subscriber \cite{VeHeDePuLaJoCoMaPi11}, PON can be one possible implementation of the optical fronthaul in L-DAS.}


\begin{figure*}[!t]
\psfrag{d}[cc][cc][1][0]{\sf $\vdots$}
\psfrag{B}[cc][cc][.8][0]{\sf baseband module}
\psfrag{R}[cc][cc][.8][0]{\sf $m$th RF module at BBU}
\psfrag{b}[lc][cc][.8][0]{\sf baseband unit (BBU)}
\psfrag{K}[cc][cc][.8][0]{\sf $m$th distributed antenna (DA) port}
\psfrag{1}[cc][cc][.8][0]{\sf $1$st Ant.}
\psfrag{m}[cc][cc][.8][0]{\sf $m$th Ant.}
\psfrag{M}[cc][cc][.8][0]{\sf $M$th Ant.}
\psfrag{3}[cc][cc][.8][0]{\sf $1$}
\psfrag{4}[cc][cc][.8][0]{\sf $m$}
\psfrag{5}[cc][cc][.8][0]{\sf $M$}
\psfrag{x}[cc][cc][.8][0]{\sf fiber}
\psfrag{e}[cc][cc][.8][0]{\sf \textcolor[rgb]{0.00,0.00,1.00}{$P_{\rm sp1}$, $P_{\rm sp2}$, $P_{\rm sig}$ (TPI)}}
\psfrag{a}[cc][cc][.8][0]{\sf \textcolor[rgb]{0.00,0.00,1.00}{$P_{{\rm cc1},m}$ (TPI)}}
\psfrag{c}[cc][cc][.8][0]{\sf \textcolor[rgb]{0.00,0.00,1.00}{$P_{{\rm cc2},m}$ (TPI)}}
\psfrag{z}[cc][cc][.8][0]{\sf \textcolor[rgb]{0.00,0.00,1.00}{$P_{\rm fix}$}}
\psfrag{u}[cc][cc][.8][0]{\sf \textcolor[rgb]{0.00,0.00,1.00}{(TPI)}}
\psfrag{o}[cc][cc][.8][0]{\sf  {$P_{{\rm tx},m}$ (TPD)}}
\begin{center}
\epsfxsize=0.9\textwidth \leavevmode \epsffile{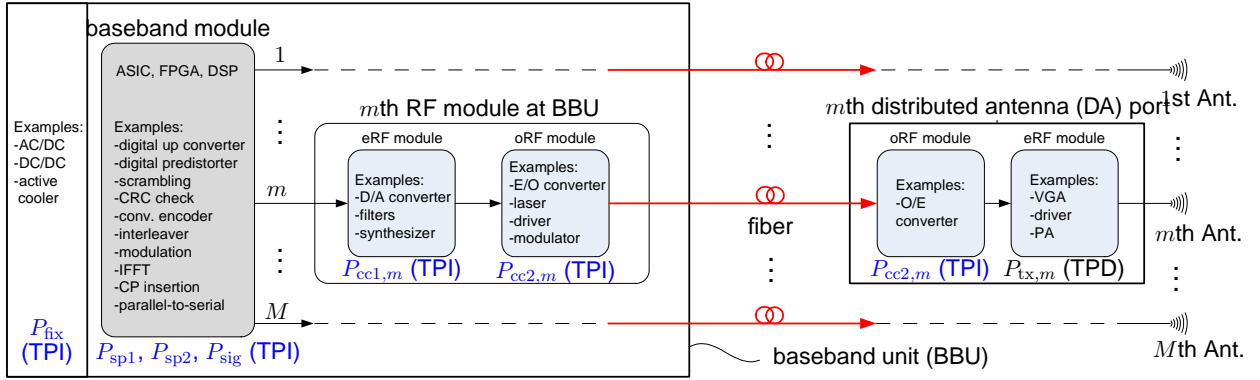}
\caption{Illustration of an L-DAS transmitter with a BBU and $M$ DAs. The BBU consists of one baseband, $M$ electric-RF (eRF), and $M$ optical-RF (oRF) modules. Each DA consists of one-pair of remote oRF and eRF modules. The TPD term $P_{{\rm tx},m}$ is the $m$th summand in (\ref{fcost}).}
\label{Fig:LDASPow}
\end{center}
\end{figure*}

Particularly, the BBU consists of a baseband module and $M$ RF modules, each of which consists of an electric RF (eRF) module and an optical RF (oRF) module as shown in Fig. \ref{Fig:LDASPow}. The baseband module performs various digital signal processing, such as digital up conversion, channel encoding, and modulation (see more examples in Fig. \ref{Fig:LDASPow}), and distributes the digital signals to $M$ eRF modules. Each eRF module converts the digital electric-signals to analogue electric-signals, which is subsequently converted to optic-signals through an electric-to-optic (E/O) converter at the oRF module. The oRF module is connected to the remote oRF module at DA port through optical fiber. Each DA port consists of the remote oRF and eRF modules. The remote oRF module converts optic-signals to electric-signals via an optic-to-electric (O/E) converter, and the eRF module simply emits the electric-signals through a variable gain amplifier (VGA), a driver, and a power amplifier (PA) followed by RF antenna. The power consumption of each module is modeled precisely in Section \ref{Sec:ProbForm}.

Since an L-DAS is characterized with the very large number of DAs, the following issues will need to be addressed.

\begin{itemize}
\item {\it Processing delay  {(computational complexity)}}: It occurs from the centralized, huge computation at the BBU. A cloud-processing at a cloudlet can support a BBU of L-DAS to relieve the high processing delay \cite{LiShJiDiYuNiLi13}.  {Through the proposed AS and UC methods in Section \ref{Sec:AntSel}, parallel processing per cluster is possible, resulting in complexity and processing delay reduction.}
\item  {{\it Fronthaul delay}: Distributing signals from BBU to DA ports experiences a fronthaul delay. The fronthaul delay is different from a backhaul delay, which is caused by the limited backhaul capacity and signaling among different cells \cite{ToPeKo11,NgLoSc12b}, as it occurs within the same cell. Significant fronthaul delay can be avoided from delay-tolerant optical fiber (around $5\us/\km$) and a residual optical delay tuner (up to $15\ns$) \cite{OkBiShZhScGaBoGa05}.}
\item {\it Propagation delay}: There is a delay in propagation of transmit signal from DA port to UE. Due to the large number of DAs distributed over the coverage area, we can almost always find the proper DA which is sufficiently close to the US such that the propagation relay is negligible. With $25$ to $900$ DAs, {\it intra-antenna distance} (IAD) that is a minimum distance of neighboring DAs varies from $30\m$ to $200\m$ (refer to Section \ref{Sec:Eval}), which is the coverage of small cells, such as pico and femto cells, where the propagation delay is not critical issues.
\item  {{\it Feedback (signaling) overhead}: For the MU-MIMO, the required amount of channel state information (CSI) is enormous due to the large number of DAs (refer to Subsection \ref{Clustering}). The burden of severe signaling overhead can be effectively reduced from the proposed AS and UC followed by the cluster-based parallel processing.}
\item {\it Energy consumption}: To activate many DAs, high energy or power consumption is expected. However, as we pointed out in Section \ref{Sec:Intro}, the energy consumption behavior of L-DAS is unclear. Throughout the paper, we focus on EE characterization of L-DAS and propose EE-improving baseband algorithms. 
\item  {{\it Remaining issues:} A synchronization issue is tightly related to the delay issues. To completely resolve the recognized delay issues and synchronization issues is out of scope of our work. Please refer to Section \ref{Sec:Conc} for other notable, remaining issues.}
\end{itemize}
 {Parts of issues of L-DAS, such as high computational complexity, resource consumption for feedback overhead, and energy consumption, have been rigorously considered in the paper.} For convenience, some notations and symbols used throughout the paper are listed in Table \ref{table0}.

\begin{table}[!t]
\centering
\caption{ {Some Notations and Symbols Used in This Paper.}}
\label{table0}
\setlength{\tabcolsep}{.4em}
\renewcommand{\arraystretch}{1}
\begin{tabular}{ c  l }\hline
$a$, ${\bm a}$, ${\bm A}$ & scalars, vectors, matrices\\
$|a|$, $\|{\bm a}\|$, $\|{\bm A}\|_{F}$ & absolute value, 2-norm, Frobenius norm{\vspace{0.3mm}}\\
$\dim({\bm A})$ & dimension of matrix ${\bm A}$\\
$\mathcal{A}$, $|\mathcal{A}|$ & sets, cardinality of set\\
$(\cdot)^T$, $(\cdot)^H$, $(\cdot)^{\dag}$ & transposition, Hermitian transposition, pseudo-inverse\\
$\tr(\cdot)$, $\E(\cdot)$ & trace of matrix, expectation of a random variable\\
${\bm a}_1\circ{\bm a}_2$, ${\bm A}_1\circ{\bm A}_2$ & an element-wise product of vectors or matrices \\
$[{\bm a}]_{m}$, $[{\bm A}]_{mu}$ & the $m$th and ($m,u$)th elements\\
${\bm a}^{\mathrm{c}}_{m}$, ${\bm a}^{\mathrm{r}}_{m}$ & the $m$th column and row vectors of a matrix ${\bm A}${\vspace{0.3mm}}\\
$(\cdot)^\prime$ and $(\cdot)^*$ & a given constant value, an optimized variable{\vspace{0.3mm}}\\\hline
$U$, $M$, $L$ & numbers of UEs, DAs, and clusters in network\\
$u$, $m$, $\ell$ & indices of UE, DA, and cluster\\
$U_\ell$, $M_\ell$ & numbers of UEs and DAs allocated to cluster $\ell$\\
$M_u$ & number of DAs assigned to UE $u$\\
$\mathcal{U}$ & set of all UEs, defined as $\{1,\ldots,U\}\ni u$\\
$\mathcal{M}$ & set of all DAs, defined as $\{1,\ldots,M\}\ni m$\\
$\mathcal{L}$ & set of clusters, defined as $\{1,\ldots,L\}\ni \ell$\\
$\mathcal{U}_\ell$, $\mathcal{M}_\ell$ & sets of UEs and DAs in cluster $\ell$ \\
$\mathcal{M}_u$ & set of DAs assigned to UE $u$\\
${\bm H}\in\mathbb{C}^{U\times M}$ & channel matrix \\
${\bm H}_\ell\in\mathbb{C}^{U_\ell\times M}$ & channel matrix for cluster $\ell$ \\
${\bm S}\in\mathbb{R}^{M\times U}$ & antenna selection matrix (binary)\\
${\bm S}_\ell\in\mathbb{R}^{M\times U_\ell}$ & antenna selection matrix for cluster $\ell$ (binary)\\
${\bm W}\in\mathbb{C}^{M\times U}$ & precoding matrix\\
${\bm W}_\ell\in\mathbb{C}^{M\times U_\ell}$ & precoding matrix for cluster $\ell$\\
${\bm P}\in\mathbb{R}^{U\times U}$ & power control matrix (diagonal)\\
${\bm P}_\ell\in\mathbb{R}^{U_\ell\times U_\ell}$ & power control matrix for cluster $\ell$ (diagonal)\\
${\bm y}\in\mathbb{C}^{U\times 1}$ & received symbol vector\\
${\bm x}\in\mathbb{C}^{U\times 1}$ & transmit symbol vector \\
${\bm n}\in\mathbb{C}^{U\times 1}$ & AWGN vector \\
$P_m$ & maximum output power of DA $m$\\
$R_u$ & target rate of UE $u$\\\hline
\end{tabular}
\end{table}


\section{EE Maximization Problem Formulation}\label{Sec:ProbForm}

For simplicity, we assume that i) each UE has a single receive antenna and ii) any channel matrix of the selected DAs is full rank. Denoting a received signal at UE $u$ by $y_u$, its vector form ${\bm y}=[y_1\cdots y_U]^T$ is written as
\beq\label{RxSig}
{\bm y} = {\bm H} \left({\bm S}\circ{\bm W}\right)\!  \sqrt{\bm P}{\bm x} + {\bm n},
\eeq
where ${\bm H}$ is a $U$-by-$M$ MU-MIMO channel matrix\footnote{ {The channels are assumed to be sufficiently static for MU-MIMO precoding, i.e., large coherence time of channels. This assumption is supported by realizing a user association or scheduling, which groups $U$ users who move with a low mobility less than $1\Kmph$ for example.}}; ${\bm S}$ is an $M$-by-$U$ binary, transmit AS matrix whose $(m,u)$th element $s_{mu}=1$ if the $m$th DA is selected for UE $u$, and $s_{mu}=0$ otherwise; ${\bm W}$ is an $M$-by-$U$ precoding matrix; ${\bm P}$ is a $U$-dimensional diagonal matrix whose $u$th diagonal element ${p_{uu}}$ determines a power portion assigned to UE $u$; ${\bm x}=[x_1\cdots x_U]^T$ is a transmit signal vector where $x_u$ is a transmit symbol to UE $u$ with $\E|x_u|^2=1$; and ${\bm n}=[n_1\cdots n_U]^T$ is an additive white Gaussian noise (AWGN) vector whose $u$th element $n_u$ is an AWGN at UE $u$ and obeys the complex normal distribution with a zero mean and a $\sigma^2$ variance, i.e., $\mathcal{CN}(0,\sigma^2)$. The $(u,m)$th element of ${\bm H}$ represents a channel gain $\sqrt{A_{um}}h_{um}$ consisting of the path loss $\sqrt{A_{um}}$ and the small scale fading $h_{um}$ between DA $m$ and UE $u$. The channels $\{h_{um}\}$ are assumed to be independent and identically distributed (i.i.d.).

The received signal-to-interference-plus-noise ratio (SINR) of UE $u$ is derived from $y_u$ in (\ref{RxSig}) as
\beqn
{\sf SINR}_u ({\bm S},{\bm W}\!,{\bm P})
&=& \frac{\E \left|{\bm h}^{\mathrm{r}}_{u} \left({\bm s}^{\mathrm{c}}_{u}\circ {\bm w}^{\mathrm{c}}_{u}\right)\sqrt{p_{uu}} x_u \right|^2}{\sum_{u'=1,u'\neq u}^U \E\left|{\bm h}^{\mathrm{r}}_{u} \left({\bm s}^{\mathrm{c}}_{u'}\circ {\bm w}^{\mathrm{c}}_{{u'}}\right)\sqrt{p_{{u'}{u'}}} x_{u'}\right|^2+ \sigma^{2}}\nonumber\\
&=&\frac{\left|{\bm h}^{\mathrm{r}}_{u} \left({\bm s}^{\mathrm{c}}_{u}\circ {\bm w}^{\mathrm{c}}_{u}\right)\right|^2 p_{uu}}{\sum_{u'=1,u'\neq u}^U \left|{\bm h}^{\mathrm{r}}_{u} \left({\bm s}^{\mathrm{c}}_{u'} \circ {\bm w}^{\mathrm{c}}_{{u'}}\right)\right|^2 p_{{u'}{u'}}+\sigma^{2}}.\label{SINR}
\eeqn
Under sufficient input backoff assumption, a PA input signal is linearly amplified and the PA output signal has a Gaussian distribution \cite{JoHoSu14JSAC}. Hence, we can further assume that UE $u$ can achieve throughput over bandwidth $\Omega\Hz$ as\footnote{ {We consider the system performance after equalization. The equalizer can be located at the receiver or the transmitter in the form of orthogonal frequency-division multiplexing (or similar techniques), and is assumed to be able to remove the effects of inter-symbol interference (ISI) over any frequency band. The ISI therefore is ignored from the performance evaluation of our system with arbitrary $\Omega$.}}
\beq\nonumber
{\sf R}_u ({\bm S},{\bm W}\!,{\bm P}) = \Omega \log_2\left(1+ {\sf SINR}_u ({\bm S},{\bm W}\!,{\bm P})\right),~\forall u\in\mathcal{U}.
\eeq
The system throughput per unit time ($\operatorname{bits/sec}$) is then written as
\beq\label{TP1}
{\sf R} ({\bm S},{\bm W}\!,{\bm P}) =  \sum_{u\in\mathcal{U}} {\sf R}_u ({\bm S},{\bm W}\!,{\bm P}).
\eeq

Now, we propose a power consumption model which coincides with signal model (\ref{RxSig}) and captures the effect of the core design factors of L-DAS, such as $U$, $M$, and maximum output power of DA. The power consumption function ${\sf C}(\cdot)$ of L-DAS transmitter is basically modeled as two parts as
%
%
\beq\label{Pc2}
{\sf C}({\bm S},{\bm W}\!,{\bm P})= f({\bm S},{\bm W}\!,{\bm P}) + g({\bm S},{\bm W}),
\eeq
where the first part $f(\cdot)$ is a transmit power dependent (TPD) term and the second part $g(\cdot)$ is a transmit power independent (TPI) term (refer to Fig. \ref{Fig:LDASPow}).

The TPD power consumption is the sum of all TPD terms of DAs as
\beq \label{fcost}
f({\bm S},{\bm W}\!,{\bm P}) = c \sum_{m\in\mathcal{M}} \frac{1}{\eta_m}\left[(\bm{S}\circ{\bm W}){\bm P}({\bm S}\circ{\bm W})^H\right]_{mm}, 
\eeq
where $c$ is a system dependent power loss coefficient ($c>1$) which can be empirically measured, $\eta_m$ is the efficiency of PA at the $m$th DA ($0<\eta_m<1$), and the $(m,m)$th element of the matrix inside a bracket is the transmit power of DA $m$. The average transmit power of DA $m$ is derived from (\ref{RxSig}) as
\beq\nonumber
\begin{split}
\E\left|\left({\bm s}^{\mathrm{r}}_{m}\circ{\bm w}^{\mathrm{r}}_{m}\right)\sqrt{{\bm P}}{\bm x}\right|^2
&=\E \left( \left({\bm s}^{\mathrm{r}}_{m}\circ{\bm w}^{\mathrm{r}}_{m}\right) \sqrt{{\bm P}}{\bm x}{\bm x}^{H}\sqrt{{\bm P}}\left({\bm s}^{\mathrm{r}}_{m} \circ {\bm w}^{\mathrm{r}}_{m}\right)^H \right)\\
&= \left({\bm s}^{\mathrm{r}}_{m}\circ{\bm w}^{\mathrm{r}}_{m}\right){\bm P}\left({\bm s}^{\mathrm{r}}_{m} \circ {\bm w}^{\mathrm{r}}_{m}\right)^H\\
&= \left[\left(\bm{S}\circ{\bm W}\right){\bm P}\left({\bm S}\circ {\bm W}\right)^H\right]_{mm}.
\end{split}
\eeq

The TPI power consumption is modeled as
\beq\label{PC-nTx}
g({\bm S},{\bm W})
= A({\bm S})+ B({\bm W}) + C(M) + \Pfix,
\eeq
where $A({\bm S})$ is the power consumption of an RF circuit, which is proportional to the number of RF chains and depends on the type of eRF and oRF modules; $B({\bm W})$ is the power consumption of signal-processing at BBU, which depends on baseband processing including precoding ${\bm W}$; $C(M)$ is the power consumption for overhead signaling which depends on network size $M$; and $P_{\rm fix}$ is the fixed power consumption including a part of power consumption at, for example, a power supply, an alternating current to direct current (AC/DC) converter, a DC/DC converter, and an active cooling system at BBU and/or DAs. We further precisely model $A({\bm S})$, $B({\bm W})$, and $C(M)$ in (\ref{PC-nTx}) as follows (refer to Fig. \ref{Fig:LDASPow}):
\beqi\IEEEyesnumber\label{PowModel}
A({\bm S})&\;=\;&\sum_{m\in\mathcal{M}}\! \big(P_{{\rm cc1},m} + P_{{\rm cc2},m} \sum_{u\in \mathcal{U}} R_u \big)\maxj_{u} s_{mu},~~~~\inum\label{Apower}\\
B({\bm W})&\;=\;&\Omega P_{\rm sp1}\left[\dim( {\bm W}) \right]^{\beta+1}  + \Omega P_{\rm sp2},\inum\label{Bpower}\\
C(M)&\;=\;&M \Omega P_{\rm sig}.\inum\label{Cpower}
\eeqi

In (\ref{Apower}), $P_{{\rm cc1},m}$ is the power consumption of eRF module at BBU for DA $m$, which includes digital-to-analogue (D/A) converter, filters, synthesizer, and mixer; $P_{{\rm cc2},m}$ is the power consumption per unit-bit-and-second of oRF modules connected the $m$th fiber line, which includes possibly modulator driver, laser, optical amplifier, and E/O and O/E converters; and $R_u$ is a target rate of UE $u$.

In (\ref{Bpower}), the first term is proportional to the number of active RF chains with order of $\beta\geq0$. The active RF-chain number is the same as the dimension (number of columns) of precoding matrix ${\bm W}$. The exponent $\beta$ implies the overhead power consumption of MU processing compared to SU processing. If $\beta=0$, there is no overhead for MU-MIMO signal processing computation\footnote{This ($\beta=0$) is the same power consumption model in \cite{JoSu13,JoChSu13} where the high power consumption for MU-MIMO precoding was not addressed. For example, if $\beta=0$ and $\dim({\bm W})=4$, there is no difference of signal processing power consumption between four-individual SU processing and single $4$-by-$4$ MU processing.}. If $\beta>0$, MU-MIMO signal processing computation consumes relatively higher power than SU signal processing. The maximum of exponent $\beta$ is assumed to be no greater than two as the computational complexity for $m$-dimensional MU-MIMO precoding, e.g., zero-forcing (ZF) MU-MIMO precoding, is roughly $\mathcal{O}(m^3)$, while that for SU is $\mathcal{O}(m)$; therefore, $0\leq \beta \leq 2$ is a reasonable assumption. In the second term of (\ref{Bpower}), $P_{\rm sp2}$ is the signal processing related power consumption per unit frequency at the baseband module, which is independent of the number of active RF chains.

In (\ref{Cpower}), $P_{\rm sig}$ is the signaling power consumption for channel estimation per frequency at the baseband module, which depends on a network size. Here, we simply model the network size as a linear function $M$, which is actually dependent on network topology. The reasonable value of $P_{\rm sig}$ is assumed to be between $0.5\%$ and $50\%$ of $P_{\rm sp1}$ (see Table \ref{table2} in Section \ref{Sec:Eval}).

From (\ref{TP1}) and (\ref{Pc2}), we express a {\it system} EE (not the sum of {\it per-user} EE) as a function of ${\bm S}$, ${\bm W}$, and ${\bm P}$ as
\beq\nonumber\label{EE}
{\sf EE}\left({\bm S},{\bm W}\!,{\bm P}\right)\triangleq \frac{{\sf R} ({\bm S},{\bm W}\!,{\bm P})}{{\sf C}\left({\bm S},{\bm W}\!,{\bm P}\right)},
\eeq
and formulate the EE maximization problem as follows:
\beqi\label{Opt:1}
&\underset{\{{\bm S},{\bm W}\!,{\bm P}\}}{\maxj}&
{\sf EE}\left({\bm S},{\bm W}\!,{\bm P}\right)
\inum\label{Obj:1}\\
&\mathsf{~s.t.~}&
\left[({\bm{S}\circ{\bm W}){\bm P}({\bm S}\circ{\bm W})^H} \right]_{mm} \le P_m, \; \forall m \in \mathcal{M},~~~~ \inum\label{const:power}\\
&&~ {\sf R}_u ({\bm S},{\bm W}\!,{\bm P}) \ge R_u, \; \forall u \in \mathcal{U}, \inum \label{const:rate}\\
&&~  p_{u_1u_2} = 0, \forall u_1 \neq u_2\in\mathcal{U},\inum\label{const:diag2}\\
&&~  s_{mu} \in \{0,1\}, \forall m \in\mathcal{M}, \forall u\in\mathcal{U},\inum\label{const:sel}
\eeqi
where (\ref{Obj:1}) is the objective function; (\ref{const:power}) follows a {\it per-antenna average power} constraint\footnote{Per-antenna {\it instantaneous} transmit power constraint is considered to avoid PA clipping effect in \cite{JoSu13}, while per-antenna {\it average} power constraint is considered in \cite{JoChSu13}. The average transmit power is also an important metric to characterize the transmitter's efficiency and used typically in transmitter design (see e.g., \cite{IsChJoFe12}).}, which is induced by different maximum power capability of the PA, denoted by $P_m$, and radio regulations; the inequalities in (\ref{const:rate}) are {\it per-user rate} constraints, i.e., quality-of-service (QoS) constraints; (\ref{const:diag2}) follows from the diagonal structure of ${\bm P}$; and (\ref{const:sel}) is for AS.

 {The problem (\ref{Opt:1}) may be infeasible as (\ref{const:power}) and (\ref{const:rate}) give the upper and lower bounds of transmit power, respectively, which may not be satisfied simultaneously for any selected DAs and precoding. In the infeasible case, two options may be possible i) to discard the outage UEs who do not achieve their own target rate, and solve again the new optimization problem with the remaining feasible UE set or ii) to reduce the target rates of the outage UEs. Note that it is difficult to immediately check the feasibility of the problem. Assigning all DAs to all UEs does not guarantee the feasibility. Rather than that, the use of all DAs may increase infeasibility because it increases the dimension of channel matrix, resulting in high probability of ill-conditioned channel matrix. The ill-conditioned channel matrix increases transmit power while the transmitter performs ZF-MU-MIMO precoding. Similar effect on L-MIMO with ZF-MU-MIMO was reported in \cite{BjSaHoDe13}. If a particular DA exceeds its maximum output power, transmit power of all DAs coupled through the ZF property should be scaled down, resulting in significant throughput degradation. This gives us a strong motivation to consider AS for the L-DAS with ZF-MU-MIMO later.}


Obtaining $\{{\bm S},{\bm W}\!,{\bm P}\}$ jointly by directly solving \eqref{Opt:1} is difficult due to the non-convex objective function and constraints\footnote{ {Contrary to the {\it power minimization} problem in \cite{ChPePh13}, which can be relaxed to second-order cone program and solved efficiently with the branch-and-cut (BnC) method, (\ref{Opt:1}) is still non-convex even after the continuous relaxation. Furthermore, since computable upper bound for problem (\ref{Opt:1}) is unavailable, BnC type of methods is not applicable to (\ref{Opt:1}).}}, and the integer optimization variables $\{s_{mu}\}$. Moreover, enormous computational complexity is expected as the network size of L-DAS, i.e., size of matrices $\{{\bm S},{\bm W}\!,{\bm P}\}$, is very large. Moreover, obtaining the full CSI, i.e., $U$-by-$M$ complex-valued matrix ${\bm H}$, at the BBU is a burden for the network and resource management. Therefore, instead of solving (\ref{Opt:1}) directly, we propose i) a suboptimal, cluster-based decomposition approach by determining ${\bm S}$ in Section \ref{Sec:AntSel}, and ii) further decomposition of the per-cluster subproblem into two optimization problems to find ${\bm W}$ and ${\bm P}$ in Sections \ref{Sec:ZF} and \ref{SEC:Pow}, respectively.

\section{Antenna Selection and User Clustering}\label{Sec:AntSel}
To resolve the complexity issues of solving \eqref{Opt:1}, we proposed the AS and UC methods which enable us to decompose \eqref{Opt:1} into multiple subproblems by a cluster. We cluster the UEs (or equivalently the selected DAs) based on an SINR threshold such that inter-cluster interferences (ICIs) are small enough to split the original optimization problem (\ref{Opt:1}) into the cluster-based subproblems. {Consequently, cluster-based parallel computation and feedback can reduce computational complexity and feedback (or signaling) information, and make L-DAS practicable.}

\subsection{Antenna Selection (AS) Algorithms}
 {As mentioned in the previous section, AS could be a crucial strategy to improve EE of the L-DAS with ZF-MU-MIMO. Moreover, our previous studies on a fundamental SE-EE tradeoff \cite{JoHoSu14JSAC} motivate us to consider an AS strategy, which can control SE-EE tradeoff and achieve the Pareto optimal tradeoff \cite{JiCi12,JoHoSu13WCL,JoSu13,JoChSu13,XuQi13}.}

There are a few heuristic algorithms for the AS, such as channel norm based (CNB) greedy, precoding norm based (PNB) greedy, and power consumption based (PCB) greedy algorithms \cite{JoChSu13}. However, the CNB-, PNB-, and PCB-greedy algorithms are irrelevant for the large-size network. The three greedy methods initially assign all DAs and discard one DA in each iteration sequentially; thus, they prohibitively require high computational complexity as $M$ or $U$ increases. Precisely, the time complexities of CNB-, PNB-, and PCB-greedy algorithms are $\mathcal{O}(MU^3)$, $\mathcal{O}(MU^3)$, and $\mathcal{O}(M^2U^3)$, respectively \cite{JoChSu13}. To circumvent the high complexity, we propose two simple-yet-effective AS algorithms that determine the set of DAs assigned to UE $u$, denoted by $\mathcal{M}_u$, and the corresponding AS matrix ${\bm S}$.

\subsubsection{Channel-Gain-Based (CGB)-Greedy AS Algorithm}
 {The CGB-greedy algorithm assigns each UE to a single DA based on the channel gain, e.g., received signal strength indicator (RSSI) used for 3GPP-LTE \cite{LTE}. In other words, a UE and a DA are paired with each other whose channel gain is as large as possible. Since the BBU is able to differentiate RSSIs received from different DAs, no additional resource is required.}

Let $M_u$ be a predetermined number of DAs which are supposed to be assigned to UE $u$. We assign DA $m$ to UE $u$ whose channel gain is the strongest, and discard the DA $m$ from the subsequent allocation procedure. If $M_u$ DAs are assigned to UE $u$, discard the UE $u$ from the subsequent allocation procedure. This allocation procedure is repeated until all UEs are discarded.

 {Note that full CSI is not required for the AS procedure}, and that the original AS is an $\mathcal{O}(2^{MU})$ combinatorial problem, yet the greedy AS algorithm requires only $\mathcal{O}(U)$ time complexity with $\mathcal{O}(MU\log (MU))$ for sorting $MU$ channel gains. Thus, the computational complexity is also reduced dramatically compared to the existing greedy algorithms. The CGB-greedy AS algorithm is summarized in Algorithm \ref{alg:AS}.

\begin{algorithm}[h]
\caption{: CGB/MDB-greedy AS algorithm}\label{alg:AS}
\algsetup{indent=1.5em,
linenosize=\footnotesize,
linenodelimiter=.}
\begin{algorithmic}[1]
\STATE  {Initial setup: $\mathcal{U}=\{1,\ldots,U\}$, $\mathcal{M}=\{1,\ldots,M\}$, $s_{mn}=0,\forall m\in\mathcal{M},~\forall u\in\mathcal{U}$, $\mathcal{M}_u=\emptyset,\forall u\in\mathcal{U}$, and given $M_u$'s.}\label{b}
\WHILE{$\mathcal{U}\neq\emptyset$}
\STATE {find $\{m^*,u^*\}=\underset{m\in\mathcal{M},u\in\mathcal{U}}{\arg {\text{\sf metric}}}$.
\\ `{\sf metric}' = $\maxj |h_{um}|$ for CGB,  $\minj d_{um}$ for MDB}\label{a}
\STATE{set $s_{m^*u^*}=1$, $\mathcal{M}=\mathcal{M}\setminus m^*$ and $\mathcal{M}_{u^*}=\mathcal{M}\cup m^*$}
\IF{$|\mathcal{M}_{u^*}|=M_{u^*}$}
\STATE{$\mathcal{U}=\mathcal{U}\setminus u^*$}
\ENDIF
\ENDWHILE
\end{algorithmic}
\end{algorithm}

\subsubsection{Minimum-Distance-Based (MDB)-Greedy AS Algorithm}
Instead of channel gains, distance information between DA $m$ and UE $u$, i.e., $d_{um}$, can be considered for the AS. This strategy is typically considered for a simple system without preprocessing at the transmitter, to reduce signaling information and backhaul overhead \cite{KiLeLeLe12}. {If the BBU has the location information of UEs with free, the MDB-greedy method can simplify the AS procedure as the BBU can omit the RSSI detection procedure. This motivates us to consider a simple MDB-greedy algorithm. However, if BBU needs to perform the localization, the signaling power consumption model in (\ref{Cpower}), which is designed for CGB-greedy algorithm, should be modified. To observe the effect of signaling power consumption on EE, we consider various values of $P_{\rm sig}$ for EE evaluation in Section \ref{Sec:Eval}.}

The basic structure of MDB-greedy algorithm is the same as the CGB-greedy algorithm except the metric at line \ref{a} in Algorithm \ref{alg:AS}.

\subsection{User Clustering (UC) Algorithm}\label{Sec:CU}

\begin{figure*}[!t]
\begin{center}\label{Fig:clustering}
\psfrag{a}[lc][cc][.6][0]{\sf \colorbox{white}{UE}}
\psfrag{c}[lc][cc][.6][0]{\sf \colorbox{white}{activated DA}}
\psfrag{e}[lc][cc][.6][0]{\sf \colorbox{white}{non-activated DA}}
\psfrag{n}[lc][cc][.6][0]{\sf \colorbox{white}{cluster with a single UE}}
\psfrag{o}[lc][cc][.6][0]{\sf \colorbox{white}{cluster with four UEs}}
\psfrag{z}[cc][cc][.6][0]{\sf \colorbox{white}{BBU}}
\psfrag{x}[lc][lc][.6][0]{\sf \colorbox{white}{IAD}}
\psfrag{u}[lc][lc][.7][0]{\sf \colorbox{white}{IAD}}
\psfrag{v}[cc][cc][.6][0]{\sf \colorbox{white}{BBU}}
\subfigure[\!\!\!\!\!]{%
\epsfxsize=0.27\textwidth \leavevmode
\epsffile{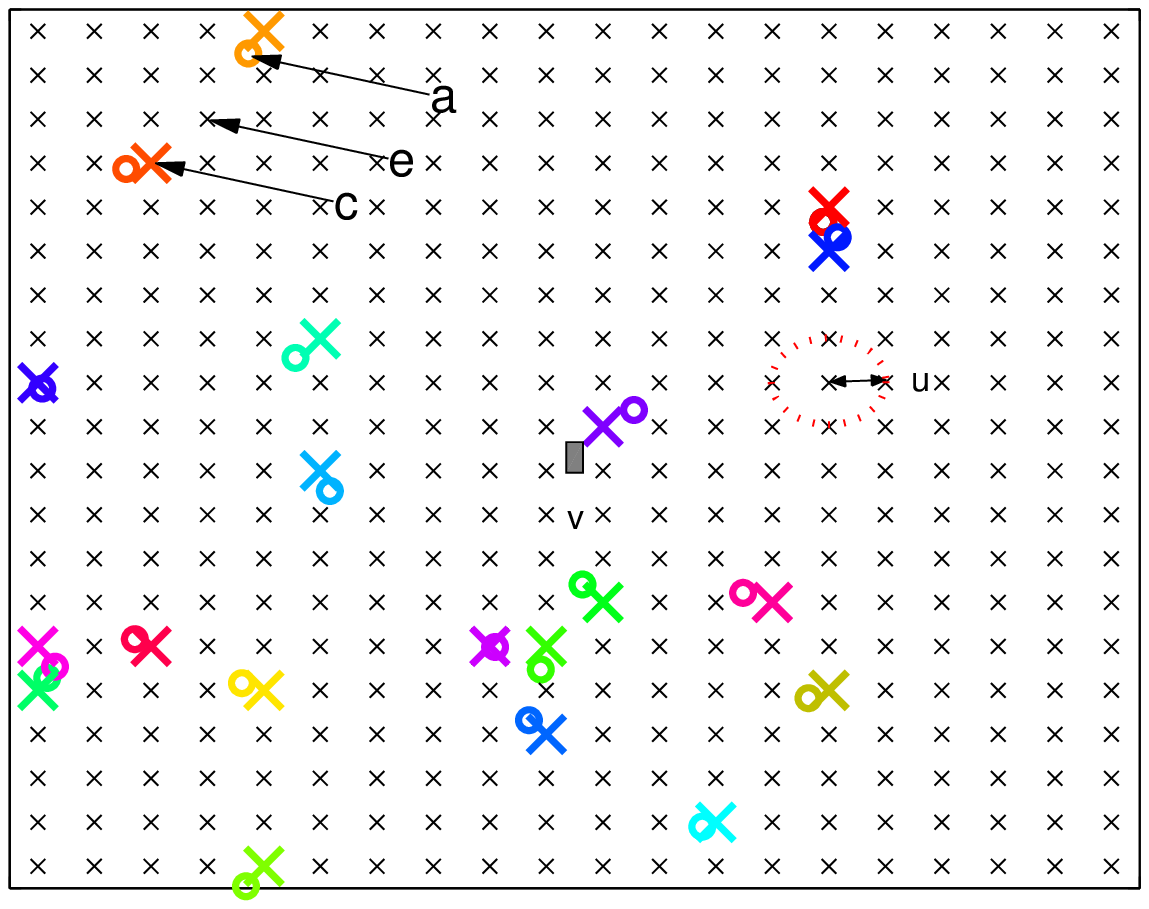}}\quad\quad
\subfigure[\!\!\!\!\!]{%
\epsfxsize=0.27\textwidth \leavevmode
\epsffile{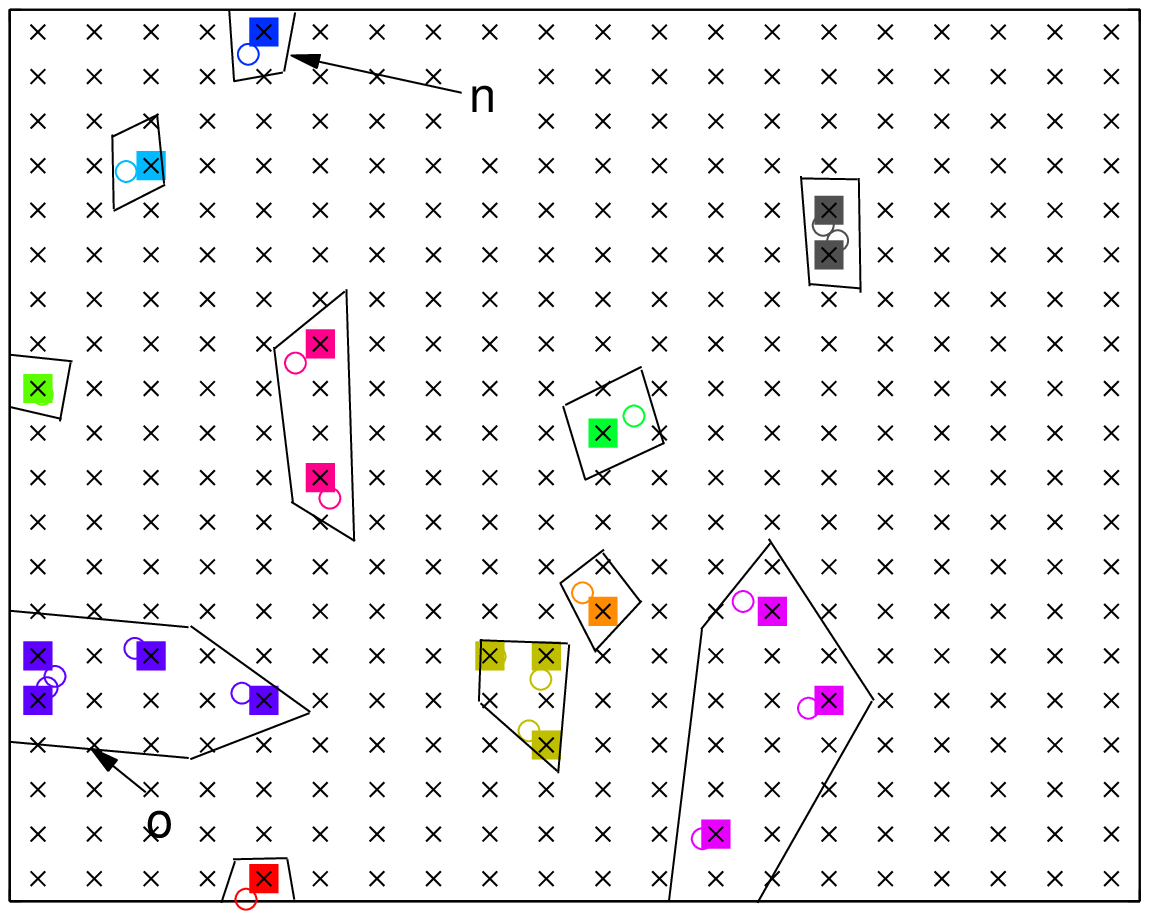}}\quad\quad
\subfigure[\!\!\!\!\!]{%
\epsfxsize=0.27\textwidth \leavevmode
\epsffile{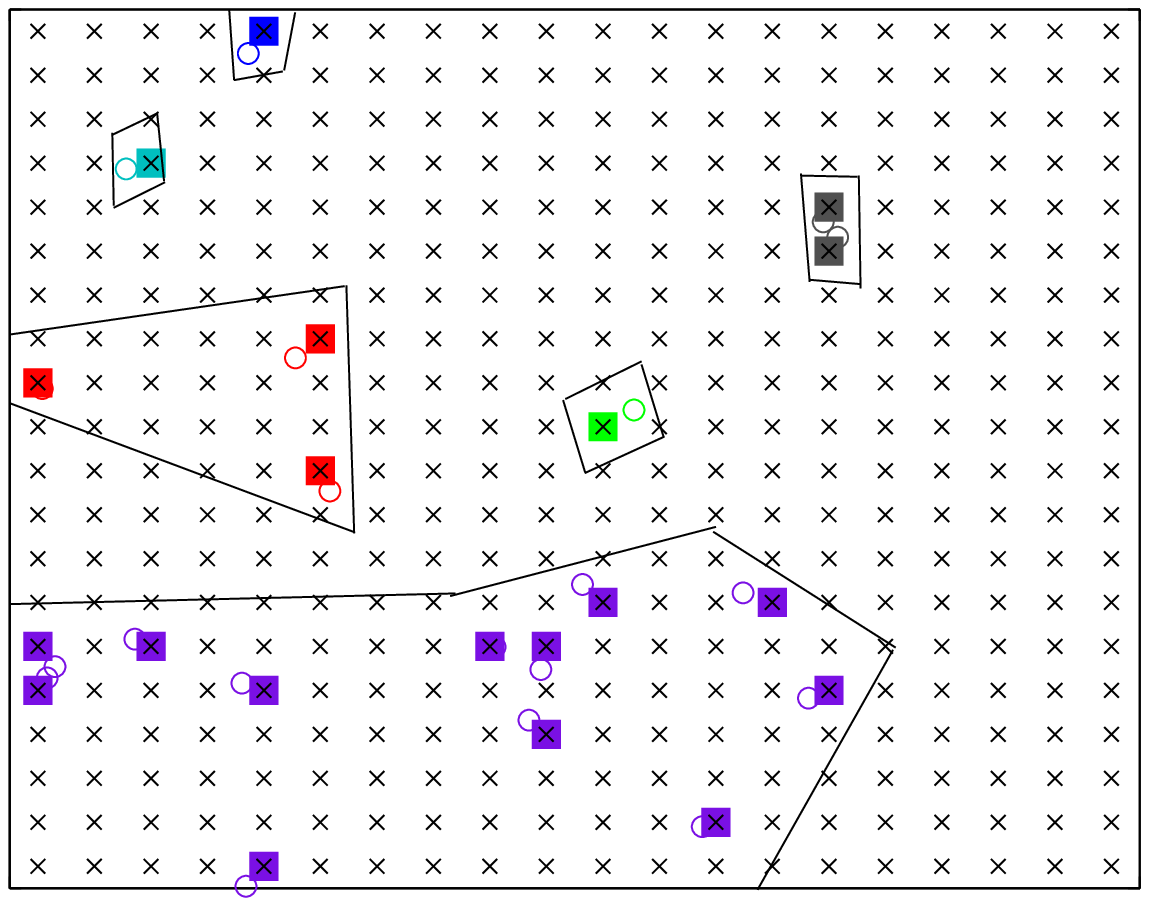}}
\caption{Illustration of AS and UC for $20$ UEs ($U=20$) and $400$ distributed antennas (DAs). a) AS through a CGB/MDB-greedy algorithm with $M_u=1,\forall u\in\mathcal{U}$. b) Clustering ($L=11$) through Algorithm \ref{alg:cluster} with $\gamma=25\dB$. c) Clustering ($L=6$) through Algorithm \ref{alg:cluster} with $\gamma=32\dB$.}
\end{center}
\end{figure*}


 {After the AS, DAs are dedicated to the specific UEs according to $\mathcal{M}_u$}, and we can then define a minimum SINR between two UEs $u$ and $u'$ as 
\beq\label{dis}
\setcounter{equation}{9}%
d(u,u') \triangleq  \minj\left\{  \frac{\sum_{m\in\mathcal{M}_u}|h_{um}|^2P_m}{\sigma^2+\sum_{m'\in\mathcal{M}_{u'}}|h_{um'}|^2P_{m'}},~ \frac{\sum_{m'\in\mathcal{M}_{u'}}|h_{u'm'}|^2P_{m'}}{\sigma^2+\sum_{m\in\mathcal{M}_u}|h_{u'm}|^2P_{m}} \right\}.
\eeq
The SINR in (\ref{dis}) is defined under the assumptions of maximum transmit power and maximum ratio combining and it can be interpreted as a distance metric between UE $u$ and $u'$ for UC. Based on the distance (not physical distance) between UEs, we cluster those that are close to each other. In other words, if two UEs are located too close to each other and the corresponding SINR is too low due to the strong inter-user-interferences (IUIs), we cluster and support them by using MU-MIMO precoding.

Denote the set of UEs in cluster $\ell$ by $\mathcal{U}_{\ell}$ such that $\bigcup_{\ell} \mathcal{U}_\ell = \mathcal{U}$ and $\mathcal{U}_{\ell} \cap \mathcal{U}_{\ell'}=\emptyset$ where $\ell\neq\ell'$ and $|\mathcal{U}_{\ell}|=U_\ell$. With a given minimum distance $\gamma$, UE $u'$ will be included to cluster $\ell$ if their distance is shorter than $\gamma$ as follows:
\beq \label{clustering}
\mathcal{U}_{\ell} = \mathcal{U}_{\ell} \cup \{u'\},\text{~if~} D(\mathcal{U}_\ell, u') \leq \gamma,
\eeq
where the distance metric between two clusters is defined as
\beq\label{Distance}
D(\mathcal{U}_\ell, \mathcal{U}_{\ell'}) \triangleq \underset{u\in\mathcal{U}_\ell,u'\in\mathcal{U}_{\ell'}}{\minj} d(u,u').
\eeq

The complexity of the UC is linear, e.g., between $\mathcal{O}(U^2)$ and $\mathcal{O}(U^3)$ for a hierarchical clustering algorithm \cite{Le92Book}. Denoting a cluster set by $\mathcal{L}=\{1,\ldots,L\}$, the proposed UC algorithm is summarized in Algorithm \ref{alg:cluster}.

\begin{algorithm}[h]
\caption{: User clustering (UC) algorithm}\label{alg:cluster}
\algsetup{indent=1.5em,
linenosize=\footnotesize,
linenodelimiter=.}
\begin{algorithmic}[1]
\STATE  {Initial setup: distance $d_o=0$, clusters $\mathcal{U}_\ell=\{\ell\}$ where $\ell\in\mathcal{L}=\{1,\ldots,U\}$, and given $\gamma$.}
\WHILE{$d_o < \gamma$}
\STATE {find the distance of most closest pair of clusters $\mathcal{U}_\ell$ and $\mathcal{U}_{\ell'}$, i.e., $d_o= \minj_{\ell,\ell'\in\mathcal{L}, \ell\neq\ell'} D(\mathcal{U}_\ell,\mathcal{U}_{\ell'})$ in (\ref{Distance}).}
\IF{$d_o < \gamma$}
\STATE{merge clusters as $\mathcal{U}_\ell = \mathcal{U}_\ell \cup \mathcal{U}_{\ell'}$.}
\STATE{update $\mathcal{L}$.}
\ENDIF
\ENDWHILE
\end{algorithmic}
\end{algorithm}

\subsection{ {Cluster-based Subproblems}}\label{Clustering}

 {Once we complete the UC with $\gamma$, the ICIs are suppressed such that the minimum SINR in (\ref{dis}) is greater than $\gamma$. Assuming sufficiently high $\gamma$ and the correspondingly negligible ICIs\footnote{ {For example, there are no ICIs if $\gamma=\infty$, as there is only one cluster. In the case, supporting all UEs with MU-MIMO through the DAs in the cluster yields different (long) propagation delays and synchronization issues. However, with a typical $\gamma$ between $20\dB$ and $30\dB$, both assumptions of negligible ICI and short propagation hold, because i) neighboring UEs consist a cluster, ii) the clusters are separated generally far way from one another for the high EE as shown in our numerical results in Section \ref{Sec:Eval}, and iii) there is no joint beamforming (i.e., MU-MIMO precoding) among the clusters.}}, the SINR (\ref{SINR}) of UE $u$ in a cell is rewritten with only IUIs within the cluster $\ell$ as
\beq\label{SINR-cluster}
{\sf SINR}_u^\ell  ({\bm S}_\ell,{\bm W}_\ell,{\bm P}_\ell)
= \frac{\left|{\bm h}^{\mathrm{r}}_{u} \left({\bm s}^{\mathrm{c}}_{u}\circ {\bm w}^{\mathrm{c}}_{u}\right)\right|^2 p_{uu}}{\underset{u'\in\mathcal{U}_\ell,u'\neq u}{\sum} \!\!\!\!\!\left|{\bm h}^{\mathrm{r}}_{u} \left({\bm s}^{\mathrm{c}}_{u'} \circ {\bm w}^{\mathrm{c}}_{{u'}}\right)\right|^2 p_{{u'}{u'}}+\sigma^{2}},
\eeq
where ${\bm S}_\ell\in\mathbb{R}^{M\times U_\ell}$, ${\bm W}_\ell\in\mathbb{C}^{M\times U_\ell}$, and ${\bm P}_\ell\in\mathbb{R}^{U_\ell\times U_\ell}$ are per-cluster AS, precoding, and power control matrices, respectively. ${\bm S}_\ell$ and ${\bm W}_\ell$ consist of column vectors ${\bm s}^{\mathrm{c}}_{u}$ and ${\bm w}^{\mathrm{c}}_{u}$, respectively, where $u\in\mathcal{U}_\ell$. ${\bm P}_\ell$ is a diagonal matrix whose diagonal elements consist of ${p_{uu}}$, where $u\in\mathcal{U}_\ell$. Hence, the per-cluster EE is defined accordingly as
\beq\label{EEcluster}
     {\sf EE}\left({\bm S}_\ell,{\bm W}_\ell,{\bm P}_\ell\right)  \triangleq {{\sf R} ({\bm S}_\ell,{\bm W}_\ell,{\bm P}_\ell)}\big/{{\sf C}\left({\bm S}_\ell,{\bm W}_\ell,{\bm P}_\ell\right)},
\eeq
where, from (\ref{TP1})--(\ref{PowModel}),
\beq\begin{split}
     {\sf R} ({\bm S}_\ell,{\bm W}_\ell,{\bm P}_\ell)
     & = \sum_{u\in\mathcal{U}_\ell}  \Omega \log_2\left(1+ {\sf SINR}_u^\ell ({\bm S}_\ell,{\bm W}_\ell,{\bm P}_\ell)\right)\\
     {\sf C}\left({\bm S}_\ell,{\bm W}_\ell,{\bm P}_\ell\right)
     &=      c \sum_{m\in\mathcal{M}_\ell} \frac{1}{\eta_m}\left[(\bm{S}_\ell\circ{\bm W}_\ell){\bm P}_\ell({\bm S}_\ell\circ{\bm W}_\ell)^H\right]_{mm}\\
     &\quad+ \!\!\!\sum_{m\in\mathcal{M}_\ell}\!\!\!\! \big(P_{{\rm cc1},m} \!+\! P_{{\rm cc2},m} \!\sum_{u\in \mathcal{U}_\ell} \!\!R_u \big)\maxj_{u\in\mathcal{U}_\ell} s_{mu}\nonumber\\
     &\quad\quad+
     \Omega P_{\rm sp1}\left[\dim( {\bm W}_\ell) \right]^{\beta+1}  + M_\ell\Omega P_{\rm sig}\\
     &\quad\quad\quad+ \Omega P_{\rm sp2}/L + \Pfix /L.\nonumber
\end{split}
\eeq
Here, $M_\ell$ is the number of DAs assigned in cluster $\ell$ and $\mathcal{M}_\ell$ is the set of $M_\ell$ DAs assigned to cluster $\ell$.}

Using ${\bm S}_\ell^*$ obtained from AS to (\ref{EEcluster}), we split the original problem into multiple subproblems by the cluster $\ell\in\mathcal{L}$ as
\beqi\label{Opt:1s}
&\underset{\{{\bm P}_\ell,{\bm W}_\ell\}}{\maxj}&
{\sf EE}\left({\bm S}_\ell^*,{\bm W}_\ell,{\bm P}_\ell\right)
\inum\label{Obj:1s}\\
&\mathsf{~s.t.}&
\!\left[({\bm{S}_\ell^*\!\circ\!{\bm W}_\ell){\bm P}_\ell({\bm S}_\ell^*\!\circ\!{\bm W}_\ell)^H} \right]_{mm} \!\le P_m, \!\forall m \!\in\! \mathcal{M}_\ell,~~~~ \inum\label{const:powers}\\
&& {\sf R}_u (\bm{S}_\ell^*,{\bm W}_\ell,{\bm P}_\ell) \ge R_u, \; \forall u \in \mathcal{U}_\ell, \inum \label{const:rates}\\
&&  p_{u_1u_2} = 0, \forall u_1 \neq u_2\in\mathcal{U}_\ell.\inum\label{const:diag2s}
\eeqi
Since the subproblem (\ref{Opt:1s}) can be solve in parallel over $\ell$, the computational complexity of the original problem can be significantly reduced.  {Furthermore, since ${\bm h}^{\mathrm{r}}_{u} \left({\bm s}^{\mathrm{c}}_{u}\circ {\bm w}^{\mathrm{c}}_{u}\right) = \sum_{m\in\mathcal{M}_\ell} [{\bm h}^{\mathrm{r}}_{u}]_m  [{\bm w}^{\mathrm{c}}_{u}]_m$ in (\ref{SINR-cluster}), the required CSI for cluster $\ell$ is the cluster-based channel matrix instead of the full channel matrix ${\bm H}$. Let ${\bm H}_\ell\in\mathbb{C}^{U_\ell \times M}$ be a channel matrix of cluster $\ell$ that consists of row vectors ${\bm h}^{\mathrm{r}}_{u}$, $u\in\mathcal{U}_\ell$. Then, precisely, the required CSI information to solve (\ref{Opt:1s}) is only the $m$th columns of ${\bm H}_\ell$, where $m\in\mathcal{M}_\ell$. In other words, we need only $M_\ell U_\ell$ complex values for the cluster $\ell$. The remaining values in other columns of ${\bm H}_\ell$ can be set to be zeros as they will be discarded by AS matrix ${\bm S}_\ell$ during the optimization. Hence, the feedback (or signaling) information can be significantly reduced from $MU$ complex values for ${\bm H}$ to $\sum_{\ell} M_\ell U_\ell$ complex values for $\{{\bm H}_1,\ldots,{\bm H}_L\}$. In the example of Figs. 2(b) and (c), $48$ and $160$ complex values are required for the CSIs, respectively, which is huge reduction compared with $8,000$ complex values for a naive system without AS and UC.}


 {From Jensen's inequality, ${\sf EE}\left({\bm S}^*,{\bm W},{\bm P}\right) \leq \sum_{\ell} {\sf EE}\left({\bm S}_\ell^*,{\bm W}_\ell,{\bm P}_\ell\right)$. Thus, optimality loss arises from the EE upper bound maximization in (\ref{Opt:1s}). However, the decomposition strategy provides reasonable performance if the bound is sufficiently tight, and the tightness depends on $\gamma$ as follows. As the clustering threshold $\gamma$ increases, cluster size increases and the number of clusters decreases to one (see Fig. 2). In the case, (\ref{Obj:1s}) is identical to the original objective function (\ref{Obj:1}) without any optimality loss. On the other hand, as $\gamma$ decreases, the number of clusters increases resulting in optimality loss. To mitigate the optimality loss, we consider adaptive algorithms for the number of assigned DAs to each UE $M_u$ and the clustering threshold $\gamma$.}

\subsection{Adaptive Algorithms for $M_u$ and $\gamma$}\label{Sec:option}
To mitigate the optimality loss and avoid the outages, it is worth exploring additional DA assignment to the outage UE if there are unallocated DAs in a network, i.e., $\sum_{u\in\mathcal{U}} M_u < M$. For the additional DA assignment, we increase $M_u$ at line \ref{b} of Algorithm \ref{alg:AS}. With a limit of iterations, denoted by $Q_{AS}$, the AS adaptation algorithm is summarized in Algorithm \ref{alg:AS_adapt}. In Algorithm \ref{alg:AS_adapt}, precoding and power control in lines \ref{c} and \ref{d} will be introduced in the subsequent sections.

\begin{algorithm}[h]
\caption{: $M_u$ adaptation algorithm for AS}\label{alg:AS_adapt}
\algsetup{indent=1.5em,
linenosize=\footnotesize,
linenodelimiter=.}
\begin{algorithmic}[1]
\STATE{Initial setup: $M_u=1,\forall u \in \mathcal{U}$, $q=0$, $\gamma$, and $Q_{AS}\geq0$.}
\WHILE{{\sf feasibility}$=0$ \& $q < Q_{AS}$}
\STATE{AS: Algorithm \ref{alg:AS}.\label{AS}}
\STATE{UC: Algorithm \ref{alg:cluster} with a threshold $\gamma$.}
\STATE{{\sf feasibility}$=1$}
\FOR{cluster $\ell=1,\ldots,L$}
\STATE{precoding: (\ref{OptW}).}\label{c}
\STATE{power control: Algorithm \ref{alg:Bisection} or (\ref{HeuristicP}).}\label{d}
\STATE{{\bf if} power control is infeasible $\&$ $\sum_{u\in\mathcal{U}}M_u<M$}
\STATE{{\bf then} add one additional DA to UE $u\in\mathcal{U}_\ell$ who has the weakest channel gain, i.e., $M_u=M_u+1$ where $u\in\mathcal{U}_\ell$ s.t., $u=\underset{u\in \mathcal{U}_\ell}{\arg\minj} |h_{um}|$.}
\STATE{{\sf feasibility}$=0$ {\bf end if}}
\ENDFOR
\STATE{$q=q+1$}
\ENDWHILE
\end{algorithmic}
\end{algorithm}

The clustering threshold $\gamma$ in (\ref{clustering}) can be also adjusted to avoid outage or to further improve EE.  {If we increase $\gamma$, the cluster size will increase, while the number of clusters will decrease. Accordingly, throughput increases due to the reduced actual ICIs after precoding and power allocation in lines \ref{c} and \ref{d}, respectively, while the processing complexity will increase due to the enlarged cluster size, i.e., MU-MIMO matrix size (see example in Fig. 2). On the other hand, if we decrease $\gamma$, the cluster size will decrease, while the number of clusters will increase. Accordingly, the processing complexity can be decreased due to the parallel processing with small MU or SU matrices, while throughput may decrease due to the increased actual ICIs.} Therefore, there exists an optimal threshold $\gamma$ for UC. If the network has sufficiently high capability to adapt $\gamma$, the local optimal $\gamma$ can be found numerically by using for example an one-dimensional line search in Algorithm \ref{alg:gamma} and a bisection search. {From the numerical results in Section \ref{Sec:Eval}, we observe the existence of optimal $\gamma$ and the adaptation of $\gamma$ can manage the actual ICIs to improve EE.} Depending on the network requirement of the computational complexity and latency, maximum number of adaptations $Q_C$ will be limited.

\begin{algorithm}[h]
\caption{: $\gamma$ adaptation algorithm for UC}\label{alg:gamma}
\algsetup{indent=1.5em,
linenosize=\footnotesize,
linenodelimiter=.}
\begin{algorithmic}[1]
\STATE{Initial setup: $\gamma$, $\delta>0$, {\sf stop}$=0$, $q=0$, and $Q_{UC}\geq0$.}
\STATE{compute ${\sf EE}_p$ with $\gamma$: Algorithm \ref{alg:AS_adapt}.}
\STATE{compute ${\sf EE}_c$ with $\gamma=\gamma+\delta$: Algorithm \ref{alg:AS_adapt}.}
\IF{${\sf EE}_c>{\sf EE}_p$}
\STATE{${\sf EE}_p={\sf EE}_c$ and $\xi=1$.}
\ELSE
\STATE{compute ${\sf EE}_c$ with $\gamma=\gamma-2\delta$: Algorithm \ref{alg:AS_adapt}.}
\STATE{{\bf if} ${\sf EE}_c>{\sf EE}_p$ {\bf{then}} ${\sf EE}_p={\sf EE}_c$ and $\xi=-1$.}
\STATE{{\bf else} {\sf stop}$=1$ {\bf end if}}
\ENDIF
\WHILE{{\sf stop}$=0$ $\&$ $q<Q_C$}
\STATE{compute ${\sf EE}_c$ with $\gamma=\gamma+\xi \delta$: Algorithm \ref{alg:AS_adapt}.}
\STATE{{\bf if } ${\sf EE}_c>{\sf EE}_p$ {\bf then} ${\sf EE}_p={\sf EE}_c$.}
\STATE{{\bf else} {\sf stop}$=1$ {\bf end if}}
\STATE{$q=q+1$}
\ENDWHILE
\end{algorithmic}
\end{algorithm}

\section{Precoding Design}\label{Sec:ZF}
Multiple UEs in a cluster are supported by MU-MIMO precoding to overcome strong IUIs. Note that a single UE in a cluster is a special case of the MU scenario. We assume that ${\bm P}_\ell'$ will be designed to satisfy (\ref{const:powers})--(\ref{const:diag2s}) for given ${\bm S}_\ell^*$ from AS. Then, the $\ell$th subproblem (\ref{Opt:1s}) is simply rewritten as
\beq\label{Opt:W}
{\bm W}_\ell^* = \underset{{\bm W}_\ell}{\maxj}\;{\sf EE}({\bm S}_\ell^*,{\bm W}_\ell,{\bm P}_\ell^\prime).
\eeq

Since a ZF-based precoding is near optimal with respect to the SE if the signal-to-noise ratio (SNR) is high enough \cite{JoKiLiJaShChChLe06,LeJi07}, it is employed for the MU-MIMO precoding of DAS \cite{YoWaShGaZhCh10}. Assuming the ZF-MU-MIMO precoding ${\bm W}_{\ell}$, all UEs in $\mathcal{U}_\ell$ share the selected DAs with one another; therefore, the AS matrix is reconstructed as $\overline{\bm S}_\ell=[{\bm s}^{\mathrm{c},*}_{\ell} \cdots {\bm s}^{\mathrm{c},*}_{\ell}]\in\mathbb{R}^{M\times U_\ell}$ where ${\bm s}^{\mathrm{c},*}_{\ell}= \sum_{u\in\mathcal{U}_\ell}{\bm s}^{\mathrm{c},*}_{u}$. Equivalently, we can write the effective channel matrix as ${\bm H}_\ell\left(\overline{\bm S}_\ell\circ{\bm W}_\ell\right) = {\bm H}_\ell{\bm S}_{\ell}^d{\bm W}_{\ell}$, where ${\bm S}_{\ell}^d=\diag({\bm s}^{\mathrm{c},*}_{\ell})\in\mathbb{R}^{M\times M}$ is a diagonal matrix whose diagonal elements are the elements of vector ${\bm s}^{\mathrm{c},*}_{\ell}$. The ZF-MU-MIMO precoding cancels perfectly IUIs in the cluster.

Following a general ZF property, the effective channel matrix ${\bm H}_\ell{\bm S}_{\ell}^d{\bm W}_{\ell}$ should be a diagonal matrix. Noting that the different values of diagonal elements can be implemented by power control $p_{uu}$, without loss of generality (w.l.o.g.), the ZF property is degenerated to ${\bm H}_\ell{\bm S}_{\ell}^d{\bm W}_{\ell} = {\bm I}_{U_\ell}$, where ${\bm I}_{a}$ is an $a$-dimensional identity matrix. Thus, we express the structure of ZF-MU-MIMO precoding matrix as
\beq\label{ZF-W}
{\bm W}_{\ell} = ({\bm{H}_\ell{\bm S}_{\ell}^d})^{\dag} + {\rm null}({\bm{H}_\ell{\bm S}_{\ell}^d}){\bm A}_{\ell},
\eeq
where ${\bm A}_{\ell}\in\mathbb{C}^{U_\ell\times U_\ell}$ is a $U_\ell$-dimensional arbitrary matrix.

 {Using the structure in (\ref{ZF-W}), we can simplify the per-cluster SINR of UE $u$ in (\ref{SINR-cluster}) to an SNR as
\beq
{\sf SNR}_u \left({\bm P}'_\ell\right)=\frac{p'_{uu}}{\sigma^{2}},~ u\in\mathcal{U}_\ell,\label{SNR}
\eeq
and obtain an optimization problem equivalent to (\ref{Opt:W}):
\beqn
{\bm W}_\ell^* &=& \underset{{\bm W}_\ell}{\maxj}\;{\sf EE}({\bm S}_\ell^*,{\bm W}_\ell,{\bm P}_\ell^\prime)\nonumber\\
&\overset{(a)}{\equiv}& \underset{{\bm W}_\ell}{\minj}\;{\sf C}({\bm S}_\ell^*,{\bm W}_\ell,{\bm P}_\ell^\prime)\nonumber\\
&\overset{(b)}{\equiv}& \underset{{\bm W}_\ell}{\minj} \; c\!\!\!\sum_{m\in\mathcal{M}_\ell} \frac{1}{\eta_m}\left[(\bm{S}_\ell^*\circ{\bm W}_\ell){\bm P}_\ell^\prime({\bm S}_\ell^*\circ{\bm W}_\ell)^H\right]_{mm}\nonumber\\
&\overset{(c)}{\equiv}& \underset{{\bm W}_\ell}{\minj}\;\sum_{m\in\mathcal{M}_\ell}\left[(\bm{S}_\ell^*\circ{\bm W}_\ell){\bm P}_\ell^\prime({\bm S}_\ell^*\circ{\bm W}_\ell)^H\right]_{mm}\nonumber\\
&=& \underset{{\bm W}_\ell}{\minj}\;\left\| {\bm S}_\ell^d {\bm W}_\ell \sqrt{{\bm P}_\ell^\prime} \right\|_F^2, \label{Opt:W1}
\eeqn
where (a) follows the fact that rate does not depends on ${\bm W}_\ell$ as the SNR (\ref{SNR}) is not a function of ${\bm W}_{\ell}$; (b) follows that ${\bm W}_\ell$ affects only on TPD term for given ${\bm S}_\ell^*$ and ${\bm P}_\ell^\prime$;  and (c) follows that equal POC is preferred for high EE as reported in \cite{JoSu13}, and ${c}/{\eta_m}$ is then a constant.}

 {Again, using (\ref{ZF-W}) to (\ref{Opt:W1}), we have an optimization problem with respect to ${\bm A}_{\ell}$ as}
\beq\label{Opt:W2}
{\bm A}_{\ell}^* = \underset{{\bm A}_{\ell}}{\minj}
\Big\|\bm{S}^d\left({\bm{H}_\ell{\bm S}_{\ell}^d}\right)^{\dag} \sqrt{{\bm P}'} + {\bm S}_\ell^d{\rm null}\left({\bm{H}_\ell{\bm S}_{\ell}^d}\right){\bm A}_{\ell}\sqrt{{\bm P}'}
\Big\|_F^2,
\eeq
where we use the property of a Frobenius norm that $\tr({{\bm A}_{\ell}{\bm A}_{\ell}^H})=\|{\bm A}_{\ell}\|_F^2$. Since the lower bound of the objective function in (\ref{Opt:W2}) is obtained when ${\bm A}_{\ell}$ is a zero matrix (refer to the Appendix in \cite{JoChSu13}), the EE-aware precoding matrix becomes a conventional ZF-MU-MIMO precoding matrix as
\beq\label{OptW}
{\bm W}_{\ell}^* = 
\left(\bm{H}_\ell\diag({\bm s}^{\mathrm{c},*}_{\ell})\right)^{\dag}.
\eeq
For an SU cluster, refer to Remark \ref{re1}.
\begin{remark}\label{re1}
There is no loss of SE optimality of cluster $\ell$ when cluster $\ell$ includes a single UE, i.e., $U_\ell=1$, because (\ref{OptW}) is an optimal beamforming for the SU cluster.
\end{remark}

\section{Power Control}\label{SEC:Pow}

We now propose a cluster-based power control method. Per-cluster optimal power control methods are proposed for SU and MU clusters. A simple heuristic power control method is also proposed for the MU cluster.

\subsection{Optimal Power Control for MU Cluster}\label{SEC:OptPow}
Consider an MU cluster $\ell$, which supports $U_\ell$ multiple UEs. For given ${\bm S}_\ell^*$ and ${\bm W}_\ell^*$, which are obtained in Sections \ref{Sec:AntSel} and \ref{Sec:ZF}, respectively, (\ref{Opt:1s}) is rewritten as
\beqi\label{Opt:P}\IEEEyessubnumber
{\bm P}_\ell^* &=&  \underset{{\bm P}_\ell}{\maxj}~
{\sf EE}({\bm S}_\ell^*,{\bm W}_\ell^*,{\bm P}_\ell)\inum\\
&\mathsf{~s.t.~}&
\left[({\bm{S}_\ell^d{\bm W}_\ell^*){\bm P}_\ell({\bm S}^d_\ell {\bm W}_\ell^*)^H} \right]_{mm} \le P_m,\forall m \in\mathcal{M}_\ell,~~~~~~\inum\label{const:power2}\\
&&~ {\sf R}_u ({\bm P}_\ell) \ge R_u, \; \forall u \in \mathcal{U}_\ell,\inum\label{const:rate2}\\
&&~  p_{u_1u_2} = 0, \forall u_1 \neq u_2\in\mathcal{U}_\ell.\inum\label{const:diag3}
\eeqi
%

By introducing an additional variable $\xi$, we rewrite \eqref{Opt:P} as
\beqi
{\bm P}_\ell^* = \underset{{\bm P}_\ell}{\maxj} \;\xi &\mathsf{~s.t.~}& {\text{(\ref{const:power2}), (\ref{const:rate2}), (\ref{const:diag3}), and }} \inum\\
&& \sum_{u\in\mathcal{U}_\ell}{\sf R}_u({\bm P}_\ell) \geq \xi {\sf C}({\bm S}_\ell^*,{\bm W}_\ell^*,{\bm P}_\ell). ~~\inum\label{new:const}
\eeqi
%
This rewriting of the optimization problem introduces an additional constraint~\eqref{new:const} to the problem. However, for fixed $\xi$, the (\ref{const:power2}) and (\ref{const:rate2}) are convex constraints and (\ref{const:diag3}) is linear constraint; therefore, the feasibility of this optimization problem can be checked through solving a \textit{convex feasibility problem} \cite{BoVa04book}. This optimization problem is therefore quasi-convex and the optimal $\xi$ can then be found through bisection and sequentially solving the convex feasibility problem at each step of the bisection. We present the bisection search in Algorithm \ref{alg:Bisection}.
\begin{algorithm}[h]
\caption{: Per-cluster optimal power control for MU}\label{alg:Bisection}
\algsetup{indent=1.5em,
linenosize=\footnotesize,
linenodelimiter=.}
\begin{algorithmic}[1]
\STATE  {setup: $\xi_{LB}= 0$, $\xi_{UB}\simeq \infty$, and a tolerance value, $\delta>0$}
\WHILE{$\xi_{UB} -  \xi_{LB}> \delta$}
\STATE {$\xi \leftarrow (\xi_{UB} -  \xi_{LB})/2$}
\STATE{Solve convex feasibility problem with constraints (\ref{const:power2}), (\ref{const:rate2}), (\ref{const:diag3}) and~\eqref{new:const}, and find (update) ${\bm P}_\ell^*$.}
\STATE{{\bf if} infeasible {\bf then} $\xi_{UB} \leftarrow \xi$}
\STATE{{\bf else} $\xi_{LB} \leftarrow \xi$ {\bf end if}}
\ENDWHILE
\STATE {${\bm P}^*_{optimal,\ell}={\bm P}_\ell^*$}
\end{algorithmic}
\end{algorithm}

 {The complexity of Algorithm \ref{alg:Bisection} for cluster $\ell$ is approximately $\mathcal{O}(M_\ell^{3.5} \log 1/\epsilon)$, where $\epsilon$ is the degree of accuracy we desire in finding the optimal energy efficiency. This complexity analysis follows from the fact that we need to solve, essentially, a semi-definite program at every iteration, which costs about $\mathcal{O}(M_\ell^{3.5})$, and the number of iterations required to get to within $\epsilon$ of the optimal energy efficiency is $\log 1/\epsilon$ \cite{BoVa04book}.} To circumvent the high complexity of per-cluster optimal power control, we consider a non-iterative power control method in the next subsection.

\subsection{Heuristic(Optimal) Power Control for MU(SU) Cluster}\label{SEC:Heuristic}
For simple closed-form solution, we modify (\ref{Opt:P}). To this end, we decompose the power control matrix as
\beq\label{Pbar}
{\bm P}_\ell=\alpha_\ell\overline{\bm P}_\ell,
\eeq
where $\alpha_\ell$ is a common power scaling factor for power limit and target rate of UEs in cluster $\ell$; $\overline{\bm P}_\ell$ is a diagonal matrix with the diagonal element $\overline{p}_{u,\ell}$; and $\overline{p}_{u,\ell}$ is the relative power portion of UE $u\in\mathcal{U}_\ell$, such that $p_{u,\ell}=\alpha_\ell \overline{p}_{u,\ell}$ and $\sum_u \overline{p}_{u,\ell} = 1$. The relative power portion factors are determined, {\it heuristically}, based on the minimum required power for the target rate as follows \cite{JoSu13}:
\beq\label{PowerPortion}
\overline{p}_{u,\ell} = \frac{\widetilde{p}_{u,\ell}}{\sum_{k\in\mathcal{U}_\ell}\widetilde{p}_{k,\ell}},~ \forall u\in\mathcal{U}_\ell,
\eeq
where $\widetilde{p}_{u,\ell}$ is the minimum required power to satisfy (\ref{const:rate}) when ZF-MU-MIMO is employed, which is derived as
\beq\nonumber\label{gam}
\widetilde{p}_{u,\ell} = \sigma^2 \left( 2^{\frac{R_u}{\Omega}}-1\right).
\eeq

Using (\ref{Pbar}) and (\ref{PowerPortion}) to the power constraint (\ref{const:power2}), we can derive the upper bound of $\alpha_\ell$ as
\beq\label{beta}
\alpha_\ell \le \minj_{m\in\mathcal{M}_\ell}\left( {P_m}\big/{\left[\bm{S}_\ell^d{\bm W}_\ell^* \overline{\bm P}_\ell({\bm S}_\ell^d{\bm W}_\ell^*)^H \right]_{mm}} \right)\triangleq \alpha_{UB,\ell}.
\eeq
On the other hand, using (\ref{Pbar}) and (\ref{PowerPortion}) to the QoS constraint (\ref{const:rate2}), we can derive the lower bound of $\alpha_\ell$ as follows:
\beq\label{alphaLB}
\alpha_\ell \geq  \frac{\sigma^2\left(2^{\frac{R_u}{\Omega}}-1\right)}{\overline{p}_{u,\ell}}=\frac{\widetilde{p}_{u,\ell}}{\overline{p}_{u,\ell}}=\sum_{u\in\mathcal{U}_\ell}\widetilde{p}_{u,\ell} \triangleq \alpha_{LB,\ell}.
\eeq
Thus, if $\alpha_\ell$ satisfies (\ref{beta}) and (\ref{alphaLB}), i.e., $\alpha_{LB,\ell}\leq \alpha_\ell \leq \alpha_{UB,\ell}$, any $\alpha_\ell\overline{\bm P}_\ell$ satisfies (\ref{const:power2}) and (\ref{const:rate2}).

Now, for simple closed-form solution of (\ref{Opt:P}), we maximize EE lower bound instead of EE. {Using (\ref{Pbar}) and an inequality that $\sum_{u\in\mathcal{U}_\ell}\log_2(f_u)\geq U_\ell\log_2(\minj_{u}(f_u))$, problem (\ref{Opt:P}) can be modified to maximize the EE lower bound, which is tight as observed in \cite{JoSu13}}, as follows:
\beqi\IEEEyesnumber\label{Opt:P2}
\alpha_\ell^*  &=&
\underset{\alpha_\ell}{\arg\maxj} \;\frac
{\Omega U_\ell \log_2\left(1+ c_{1,\ell} \alpha_\ell \right)}
{ c_{2,\ell} \alpha_\ell  + c_{3,\ell}}\inum\label{Obj3}\\
&\mathsf{~s.t.~}& \alpha_{LB,\ell} \leq \alpha_\ell \leq \alpha_{UB,\ell},\inum\label{QoSPowC3}
\eeqi
where $c_{1,\ell}\triangleq\minj_u\{\overline{p}_{u,\ell}\}\sigma^{-2}$; $c_{2,\ell}\triangleq \sum_{m\in\mathcal{M}_\ell} \frac{c}{\eta_m}[\bm{S}_\ell^d{\bm W}_\ell^*\overline{\bm P}_\ell({\bm S}_\ell^d{\bm W}_\ell^*)^H]_{mm}$; and $c_{3,\ell} = \sum_{m\in\mathcal{M}_\ell} (P_{{\rm cc1},m} + P_{{\rm cc2},m} \sum_{u\in \mathcal{U}_\ell} R_u )\maxj_{u\in\mathcal{U}_\ell} s_{mu}+\Omega P_{\rm sp1}[\dim( {\bm W}_\ell)]^{\beta+1}  + M_\ell\Omega P_{\rm sig}+ \Omega P_{\rm sp2}/L + \Pfix /L$. Note that all $c_{1,\ell}$, $c_{2,\ell}$, $c_{3,\ell}$, $\alpha_{LB,\ell}$, and $\alpha_{UB,\ell}$ in (\ref{Opt:P2}) are constant values for given ${\bm W}_\ell^*$, ${\bm S}_\ell^*$, and $\overline{\bm P}_\ell$; and the objective function (\ref{Obj3}) is a quasi-concave function over $\alpha_\ell$. Therefore, we can readily find the maximizer $\alpha_{o,\ell}$ which makes the first derivative of the objective function in (\ref{Obj3}) to zero as
\beq\nonumber
\alpha_{o,\ell} = \frac{1}{c_{1,\ell}}\left({\rm exp}\left(1+{\sf W}\left(\frac{-1}{{\rm exp}(1)}+\frac{c_{1,\ell} c_{3,\ell}}{c_{2,\ell} {\rm exp}(1)}\right)\right)-1\right),
\eeq
Considering the feasible region (\ref{QoSPowC3}), we get the optimal feasible solution of (\ref{Opt:P2}) as
\beq\nonumber\label{OptAlpha}
{\alpha}_\ell^* = \left[ \alpha_{o,\ell} \right]_{\alpha_{LB,\ell}}^{\alpha_{UB,\ell}},
\eeq
and obtain the heuristic power control matrix as
\beq\label{HeuristicP}
{\bm P}^*_{heuristic,\ell} = \alpha_\ell^* \overline{\bm P}_\ell.
\eeq

 {Since the solution in (\ref{HeuristicP}) is obtained from heuristic approach, namely EE lower bound maximization and fixed $\overline{P}_\ell$ in (\ref{PowerPortion}), it yields performance degradation compared to ${\bm P}_{optimal,\ell}^*$ in Subsection \ref{SEC:OptPow}. However, it is noticeable that the solution in (\ref{HeuristicP}) has a tractable, closed-form expression, and the performance gap is marginal as shown in the next section. Furthermore, there is no optimality loss for SU cluster as stated in Remark \ref{re2}.}
 {\begin{remark}\label{re2}
Since the EE lower bound is identical to the EE of the SU cluster and $\overline{p}_{u,\ell}=1$ in (\ref{PowerPortion}), there is no optimality loss from power control (\ref{HeuristicP}) for the SU cluster.
\end{remark}}

 {The heuristic power control method is an $\mathcal{O}(M_\ell^3)$ algorithm. This can be seen from the fact that the bottleneck procedure includes the multiplication of two $M_\ell$-dimensional matrices to find $c_{2,\ell}$ in (\ref{Opt:P2}). Hence, the highest complexity order is $\mathcal{O}(M_\ell^3)$.}

\section{Performance Evaluation and Discussion}\label{Sec:Eval}

\begin{table}[!t]
\centering
\caption{Simulation Parameters for L-DAS/L-CAS \cite{LTE,EARTH,Tu11,JoHoSu14JSAC}}
\label{table2}
\setlength{\tabcolsep}{.4em}
\renewcommand{\arraystretch}{1}
\begin{tabular}{ c  l }\hline
Cell model & square grid ($1\km^2$)\\ 
Number of DAs/CAs & $25\leq M \leq 900$ \\
Intra-antenna distance (IAD)& from $33\m$ to $200\m$\\
Number of UEs & $2\leq U \leq 20$ \\ 
UE distribution  & Uniform ($10^4-$realization)\\
Path loss model ($f_c=2\GHz$)& $A_{um}=g-128+10\log_{10}(d_{u,m}^{-\mu})$\\
feeder loss and antenna gain & $g=5\dB$\\
Path loss exponent & $\mu=3.76$\\
Small scale fading & $h_{um}\sim\mathcal{CN}(0,1)$\\
Bandwidth & $\Omega=10\MHz$ \\
Target rate & $R_u=10\Mb$\\
Maximum Tx power & $P_m=17\dBm$\\
AWGN standard deviation& $\sigma^2 = -174\dBm \!\!/\!\Hz$\\
Power loss coefficient & $c=2.63$\\
eRF circuit pow.cons. & $P_{{\rm cc1},m}=5.7\W$\\
oRF circuit pow.cons. & $P_{{\rm cc2},m}=0.5/0\pWpbps$\\
Fixed pow.cons. & $P_{\rm fix}=34\W$\\
Signal processing pow.cons. & $P_{\rm sp1}=0.94\times1/1.1\uW\!\!/\!\Hz$\\
                            & $P_{\rm sp2}=0.54\times1/1.1\uW\!\!/\!\Hz$\\
Signaling pow.cons./antenna & $5\leq P_{\rm sig} \leq 500\nW\!\!/\!\Hz$\\
preprocessing pow.cons. ratio & $0\leq \beta \leq 2$ \\
PA efficiency  &  $\eta_m=0.08/0.6$\\
Clustering threshold & $-\infty\leq\gamma\leq \infty\dB$ \\\hline
\end{tabular}
\end{table}

Computer simulations are conducted to examine the average EE performance of the proposed L-DAS. Since an instantaneous EE is set to be zero when an outage happens, the outage performance is already involved in the average EE performance. The EE performance depends highly on the power consumption models, i.e., the power consumption of TPD and TPI terms in (\ref{fcost}) and (\ref{PC-nTx}), respectively. The simulation is performed under a typical L-DAS scenario, in which the TPI power consumption is dominant compared to the TPD power consumption due to the low power transmission of DAs. With $c=2.63$, the TPD power consumption is observed to be the portion of the TPI power consumption lower than $3\%$ in our simulation. The PA efficiency\footnote{High input backoff is desired to avoid nonlinearity at the PA, around $12\dB$, because a sophisticated, complex linearization method, e.g., predistortion, is not available the simple DA port. Hence, the PA efficiency is very low.} of all DAs is set by $8\%$, i.e., $\eta_m=0.08$, at the maximum transmit power $17\dBm$. The BBU performs the centralized, complex processing as a macro BS, yet DA port covers small areas like the small BSs. Hence, we follow macro BS's power consumption model for BBU, while follow a small cell BS, such as pico and femto BSs, for DA's power consumption model. Refer to Table \ref{table2} for other detailed parameters, which are obtained from recent studies \cite{LTE,EARTH,Tu11,JoHoSu14JSAC}. Note that providing the actual, accurate measurement of the parameters is out of scope of our work.

\subsection{Average EE over Clustering Threshold}
\begin{figure}[!t]
\psfrag{x}[cc][cc][.8][0]{\sf Clustering threshold $\gamma\dB$}
\psfrag{y}[cc][cc][.8][0]{\sf Average energy efficiency $\MbpJ$}
\psfrag{c}[lc][cc][.8][0]{\sf optimal power control}
\psfrag{e}[lc][cc][.8][0]{\sf heuristic power control}
\psfrag{v}[lc][cc][.8][0]{\sf L-CAS with $\beta=0.2$}
\psfrag{n}[lc][cc][.8][0]{\sf $\beta=0.2$}
\psfrag{m}[lc][cc][.8][0]{\sf $\beta=0.5$}
\psfrag{o}[lc][cc][.8][0]{\sf $\beta=1$}
\psfrag{z}[cc][cc][.8][0]{\sf $-\infty$}
\psfrag{u}[cc][cc][.8][0]{\sf $\infty$}
\psfrag{a}[lc][cc][.8][90]{\sf full SU, $U$ clusters}
\psfrag{s}[cc][cr][.8][90]{\sf full MU, single cluster}
\begin{center}
\epsfxsize=0.7\textwidth \leavevmode
\epsffile{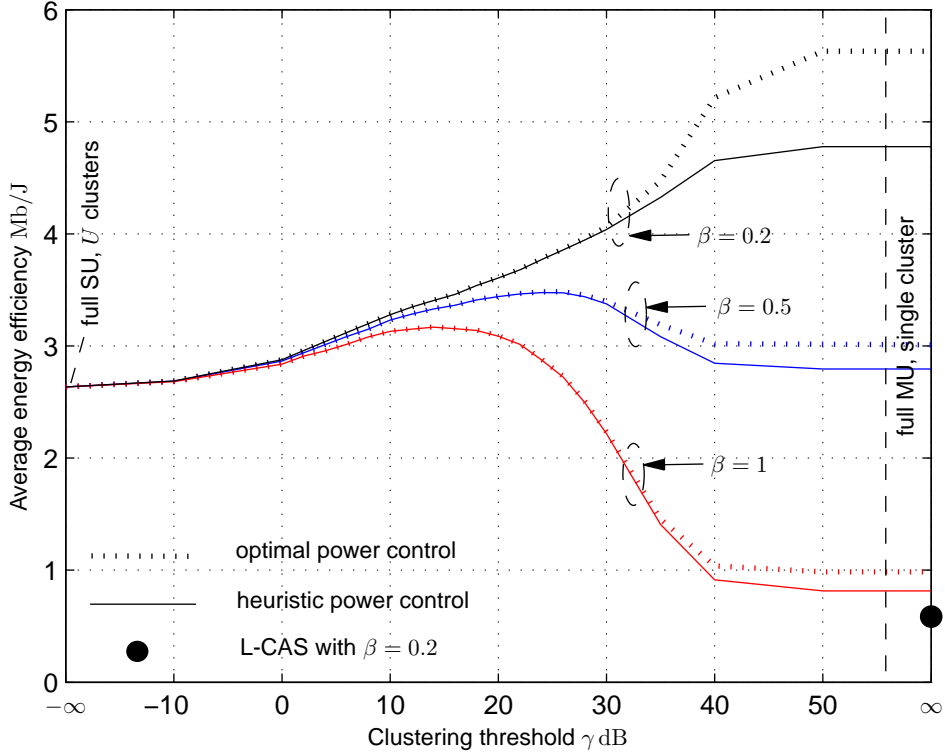}
\caption{Average EE over clustering threshold $\gamma$ with different power consumption models when $M=400$, $U=20$, and $P_{\rm sig}=50\nW\!\!/\!\Hz$.}
\label{Fig:gamma}
\end{center}
\end{figure}

Fig. \ref{Fig:gamma} shows average EEs over clustering threshold $\gamma$ with different MU-MIMO-processing power consumption models, i.e., $\beta$, when $M=400$ and $U=20$. One extreme case with $\gamma=-\infty$ yields $U$ SU clusters, each of which includes an SU; thus, it is called a {\it full SU} scheme. Other extreme case with $\gamma=\infty$ yields a single MU cluster including all UEs who are supported by ZF-MU-MIMO simultaneously, which is called a {\it full MU} scheme. If there is less penalty for MU precoding computation compared to SU precoding computation, e.g., $\beta=0.2$, EE increases as $\gamma$ increases up to the saturation of EE when $\gamma=\infty$. In other words, the full MU scheme achieves the highest EE. This is because MU-MIMO achieves the higher throughput than interference-limited SU scheme and there is small additional power consumption for MU signal processing. Note that ICI is inversely proportional to the threshold.

On the other hand, if the power consumption penalty for MU precoding increases, e.g., $\beta=1$, the overhead MU-MIMO-processing power consumption significantly decreases the EE. We observe that there exists the optimal $\gamma^*$ and the EE turns to decrease if $\gamma>\gamma^*$. In other words, the EE can be severely reduced if there are too many MU clusters in the network. As $\beta$ increases, MU signal processing has more penalty on MU-MIMO-processing power consumption, and $\gamma^*$ decreases, resulting in more SU clusters for high EE. Note that the $\gamma^*$ depends on various parameters, such as $U$, $M$, and the power consumption model, and thus, it is difficult to be found analytically. An EE gap between per-cluster optimal and heuristic power control is negligible in the contiguous the maximum EE point when the penalty of power consumption for MU precoding is nonnegligible.

 {For the sake of comparison, we add the average EE of a large-size colocated antenna system (L-CAS). The L-CAS can be interpreted as one naive implementation of L-MIMO system that employs a full MU with a simple AS method that selects $U$ antennas which give the largest average channel gains. The L-CAS may have sufficiently powerful processor to compensate nonlinear effects at the PAs; therefore, the L-CAS BS can employ a PA with much higher efficiency than the DAs in L-DAS. In simulation, we set the PA efficiency of L-CAS by $60\%$, i.e., $\eta_m=0.6$, with the corresponding signal processing power consumption increased by $10\%$. Furthermore, we set $P_{{\rm cc2},m}=0\pWpbps$ as there is no oRFs.} Numerical result shows that EE performance of L-CAS is very poor because of the high power consumption to overcome the large path losses. 

\subsection{Average EE over Number of Users}
\begin{figure}[!t]
\psfrag{x}[cc][cc][.8][0]{\sf Number of UEs, $U$}
\psfrag{y}[cc][cc][.8][0]{\sf Average energy efficiency $\MbpJ$}
\psfrag{c}[cc][cc][.8][0]{\sf optimal pow ctrl with adaptation}
\psfrag{e}[lc][cc][.8][0]{\sf heuristic pow ctrl with adaptation}
\psfrag{n}[lc][cc][.8][0]{\sf $\gamma=\infty\dB$ (full MU, single cluster)}
\psfrag{o}[lc][cc][.8][0]{\sf $\gamma=-\infty\dB$ (full SU, $U$ clusters)}
\psfrag{m}[lc][cc][.8][0]{\sf $\gamma=22\dB$}
\psfrag{v}[lc][cc][.8][0]{\sf L-CAS with $\beta=0.5$}
\begin{center}
\epsfxsize=0.7\textwidth \leavevmode
\epsffile{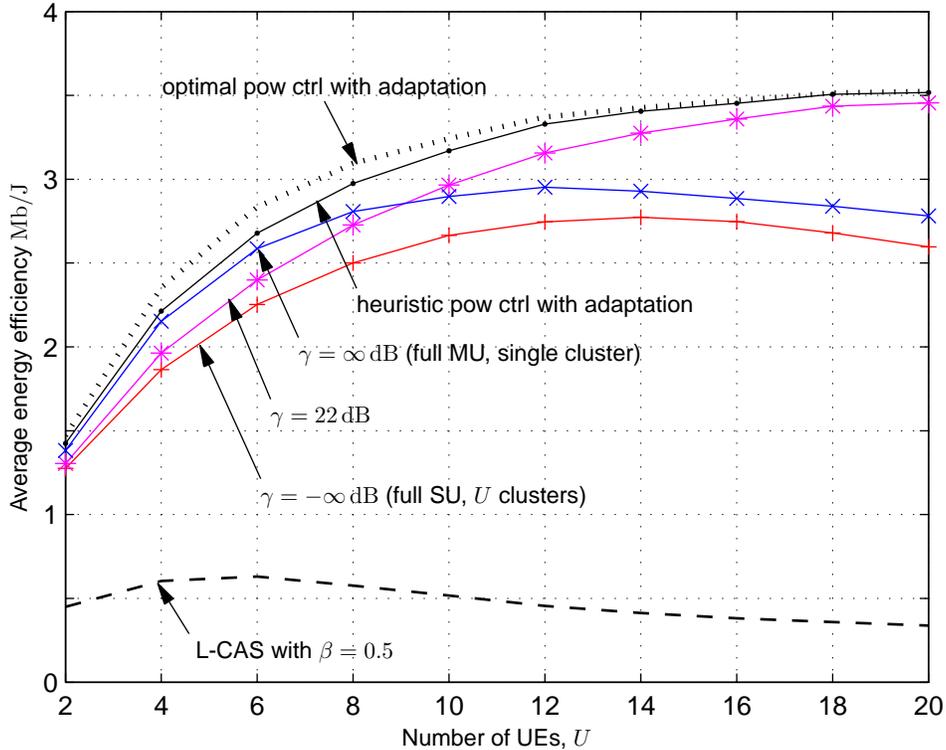}
\caption{Average EE over number of UEs, $U$, with different threshold for UC when $M=400$, $\beta=0.5$, and $P_{\rm sig}=50\nW\!\!/\!\Hz$.}
\label{Fig:U}
\end{center}
\end{figure}

Fig. \ref{Fig:U} shows average EEs over $U$ with different threshold for UC when $M=400$ and $\beta=0.5$. Two extreme cases of UC schemes with $\gamma=\infty$ and $\gamma=-\infty$ are compared with a UC with fixed $\gamma$ by $22\dB$, which allows both MU and SU schemes and is a proper threshold based on the results in Fig. \ref{Fig:gamma}. The full MU scheme achieves higher EE than the full SU scheme with heuristic power control, because the MU-MIMO-processing power consumption is not dominant when $\beta=0.5$ and the full MU-MIMO achieves higher throughput than the full SU case. As $U$ increases, UC with $\gamma=22\dB$ achieves higher EE than full MU scheme because the MU-MIMO-processing power consumption increases severely and becomes dominant, which is noticeable as considering many UEs is our interest of this work. However, when there are small number of UEs, less than nine, full MU scheme outperforms the system with $\gamma=22\dB$.

 {Using Algorithms \ref{alg:AS_adapt} and \ref{alg:gamma} in Subsection \ref{Sec:option}, $M_u$ and $\gamma$ can be adapted and EE can be improved over any number of UEs. For the adaptation parameters, we use $Q_{AS}=5$, $Q_{UC}=10$, $\delta=5\dB$, and the initial threshold $\gamma=-10\dB$. Per-cluster optimal power control further improves the EE.}


\subsection{Average EE over Network Size}
\begin{figure}[!t]
\psfrag{x}[cc][cc][.8][0]{\sf Number of transmit DAs, $M$}
\psfrag{y}[cc][cc][.8][0]{\sf Average energy efficiency $\MbpJ$}
\psfrag{c}[lc][cc][.8][0]{\sf optimal power control}
\psfrag{e}[lc][cc][.8][0]{\sf heuristic power control}
\psfrag{z}[lc][cc][.8][0]{\sf $P_{\rm sig}=5\nW\!\!/\!\Hz$}
\psfrag{u}[lc][cc][.8][0]{\sf $P_{\rm sig}=50\nW\!\!/\!\Hz$}
\psfrag{o}[lc][cc][.8][0]{\sf $P_{\rm sig}=500\nW\!\!/\!\Hz$}
\psfrag{v}[lc][cc][.8][0]{\sf L-CAS with $P_{\rm sig}=5\nW\!\!/\!\Hz$}
\begin{center}
\epsfxsize=0.7\textwidth \leavevmode
\epsffile{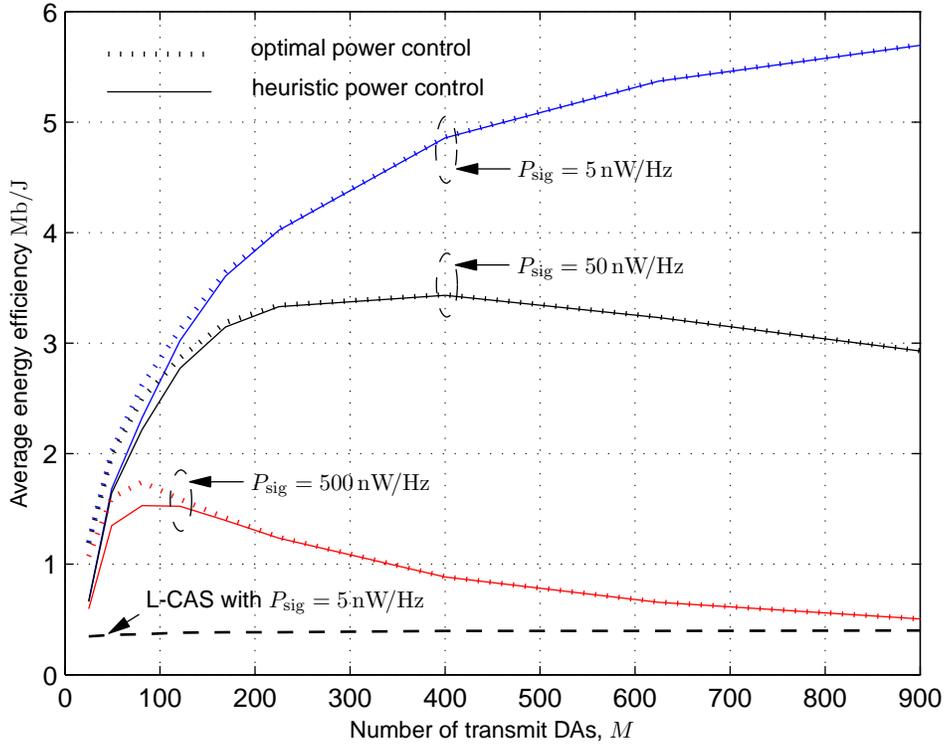}
\caption{Average EE over network size, $M$, with different threshold for UC when $U=20$, $\beta=0.5$, and $\gamma=22\dB$.}
\label{Fig:M}
\end{center}
\end{figure}

Fig. \ref{Fig:M} shows average EEs over network size $M$, i.e., the number of DAs (not active DAs), when $U=20$, $\beta=0.5$, and $\gamma=22\dB$. We evaluate EEs with three different signaling power consumption models with $P_{\rm sig}=\{5,50,500\}\nW\!\!/\!\Hz$. The average EE increases as $M$ increases because severe path loss can be circumvented with the increased degree of freedom of AS. On the other hand, the network power consumption will also increase as $M$ increases due to the nonzero $P_{\rm sig}$ even with a proper AS. Therefore, an EE increases and turns to decrease as $M$ increases, and the optimal network size is observed, e.g., around $M=400$ with $P_{\rm sig}=50\nW\!\!/\!\Hz$. As expected, the EE optimal network size decreases as $P_{\rm sig}$ increases.

\section{Conclusion}\label{Sec:Conc}
In this paper, we have considered EE maximization problem for an L-DAS. The power consumption of L-DAS transmitter has been modeled. A simple channel-gain-based antenna selection and SINR-threshold-based user clustering methods have been proposed to reduce the computational complexity of precoding and power control, and at the same time to reduce the signaling overhead. Iterative algorithms to adapt the number of assigned antennas and the clustering threshold have been considered. Numerical results have validated the potential of the L-DAS.

 {Remaining issues for further work regarding deployment, implementation, and operation of L-DAS include cell planning, regular/irregular deployment of antennas, synchronization for large cluster, robustness against CSI error, infrastructure cost for wired optical fronthaul, and a comparative, quantitative study of L-DAS and L-CAS considering capital expenditure and operational expenditure.}

\end{document}